\documentclass[aps,prl,twocolumn,showpacs,superscriptaddress,longbibliography]{revtex4-1}

\usepackage{graphicx}
\usepackage{epstopdf}
\usepackage{amsmath}
\usepackage{amsfonts}
\usepackage{amssymb}
\usepackage{color}
\usepackage{bm}
\usepackage[a4paper=true,pagebackref=false]{hyperref}
\begin{document}

\title{Experimental search for the origin of low-energy modes in topological materials}

\author{G.\,P.~Mazur} \email{grzegorz.mazur@MagTop.ifpan.edu.pl}
\affiliation{International Research Centre MagTop, Institute of Physics, Polish Academy
	of Sciences, Aleja Lotnikow 32/46, PL-02668 Warsaw, Poland}

\author{K.~Dybko}\email{dybko@ifpan.edu.pl}
\affiliation{International Research Centre MagTop, Institute of Physics, Polish Academy
	of Sciences, Aleja Lotnikow 32/46, PL-02668 Warsaw, Poland}
\affiliation{Institute of Physics, Polish Academy of Sciences, PL-02668 Warsaw, Poland}

\author{A.~Szczerbakow} \affiliation{Institute of Physics, Polish Academy of Sciences, PL-02668 Warsaw, Poland}

\author{J.\,Z.~Domagala} \affiliation{Institute of Physics, Polish Academy of Sciences, PL-02668 Warsaw, Poland}

\author{A.~Kazakov} \affiliation{International Research Centre MagTop, Institute of Physics, Polish Academy
	of Sciences, Aleja Lotnikow 32/46, PL-02668 Warsaw, Poland}

\author{M.~Zgirski} \affiliation{Institute of Physics, Polish Academy of Sciences, PL-02668 Warsaw, Poland}

\author{E.~Lusakowska} \affiliation{Institute of Physics, Polish Academy of Sciences, PL-02668 Warsaw, Poland}

\author{S.~Kret} \affiliation{Institute of Physics, Polish Academy of Sciences, PL-02668 Warsaw, Poland}

\author{J.~Korczak} \affiliation{International Research Centre MagTop, Institute of Physics, Polish Academy
	of Sciences, Aleja Lotnikow 32/46, PL-02668 Warsaw, Poland}
\affiliation{Institute of Physics, Polish Academy of Sciences, PL-02668 Warsaw, Poland}

\author{T.~Story} \affiliation{Institute of Physics, Polish Academy of Sciences, PL-02668 Warsaw, Poland}

\author{M.~Sawicki}
\affiliation{Institute of Physics, Polish Academy of Sciences, PL-02668 Warsaw, Poland}

\author{T.~Dietl} \email{dietl@MagTop.ifpan.edu.pl}
\affiliation{International Research Centre MagTop, Institute of Physics, Polish Academy
	of Sciences, Aleja Lotnikow 32/46, PL-02668 Warsaw, Poland}
\affiliation{WPI-Advanced Institute for Materials Research, Tohoku University, Sendai 980-8577, Japan}

\begin{abstract}
Point-contact spectroscopy of several non-superconducting topological materials reveals a low-temperature phase transition that is characterized by a Bardeen-Cooper-Schrieffer-type of criticality. We find such a behavior of differential conductance for topological surfaces of non-magnetic and magnetic Pb$_{1-y-x}$Sn$_y$Mn$_x$Te. We examine a possible contribution from superconducting nanoparticles, and show to what extend our data are consistent with Brzezicki's {\em et al.} theory [arXiv:1812.02168] assigning the observations to a collective state adjacent to atomic steps at topological surfaces.
\end{abstract}

\maketitle

{\em Introduction.} A series of point-contact experiments has revealed the existence of low-energy modes at junctions of metal tips with topological semiconductors and semimetals  \cite{Das:2016_APL,Aggarwal:2016_NM,Wang:2016_NM,Aggarwal:2017_NC,Wang:2017_SB,Wang:2018_SB,Naidyuk:2018_2DM,Shvetsov:2019_PRB,Zhu:2018_arXiv}. Surprisingly, despite the absence of global superconductivity in these systems, the features in d$I$/d$V$ decay critically with temperature and the magnetic field in accord with the Bardeen-Cooper-Schrieffer (BCS) theory. It has, therefore, been concluded that the superconductivity results from tip-induced strain. In the case of Cd$_{3}$As$_2$ this interpretation appears to be supported by the featureless spectrum in the case of a soft point-contact produced by silver paint \cite{Wang:2016_NM,Wang:2018_SB}. Surprisingly, however, Andreev reflection-type spectra were recently reported for Au/Cd$_{3}$As$_2$ junctions (with neither a four probe zero-resistance state nor the Meissner effect) \cite{Shvetsov:2019_PRB}, as well as for topological semimetals MoTe$_2$ \cite{Naidyuk:2018_2DM} and WC \cite{Zhu:2018_arXiv} with hard and soft point-contacts \cite{Naidyuk:2018_2DM} of various nonmagnetic and magnetic metals \cite{Zhu:2018_arXiv}. A timely question then arises about the properties of other materials, such as topological crystalline insulators (TCI) \cite{Fu:2011_PRL,Hsieh:2012_NC,Dziawa:2012_NM,Tanaka:2012_NP,Xu:2012_NC} in which hard point-contact characteristics reveal zero-bias conductance peaks (ZBCPs) \cite{Das:2016_APL}.

Here we show, employing a soft point-contact method, that a conductance gap with a broad ZBCP or Andreev-type characteristics develop at junctions of Ag with topological surfaces of diamagnetic, paramagnetic, and ferromagnetic Pb$_{1-y-x}$Sn$_y$Mn$_x$Te, where $y \gtrsim 0.67$ and $0\le x \le 0.10$, in which no signs of superconductivity are found. Nevertheless, the temperature dependence of the gap shows a BCS-like critical behavior as a function of temperature $T$ and the magnetic field $H$ with $T_c$ up to 4.5\,K and $\mu_0H_c$ up to 3\,T independent of the orientation of the magnetic field with respect to the surface plane. This implies the emergence of a collective low-temperature phase whose appearance is insensitive not only to the magnetic state of the metallic part of the junction, as found previously \cite{Das:2016_APL,Zhu:2018_arXiv}, but also to the magnetic character of the topological material. In order to elucidate the nature of these striking observations we put forward two models.

First, previous studies of the materials in question revealed superconductivity associated with metal nano-precipitates \cite{Darchuk:1998_Semicon} or misfit dislocations at the heterostructure interfaces \cite{Murase:1986_SS,Fogel:2006_PRB}. Furthermore, studies of Pb/(Pb,Sn)Te junctions point to the presence of Sn diffusion in Pb \cite{Buchner:1979_JVST}, which rises the question on whether such an effect could generate superconducting Pb or Sn at the interface.  It might be also anticipated that strain associated with different thermal expansion coefficients of silver paint and the samples or hard tips could generate misfit dislocations or precipitates. However, a series of auxiliary high-sensitivity magnetization and resistance measurements as well as high-resolution structural investigations of our samples have not revealed the presence of superconducting nanoparticles within our experimental resolution.

Second, a possible origin of a local collective phase at topological surfaces has recently been proposed by Brzezicki, Wysoki\'nski, and Hyart (BWH) \cite{Brzezicki:2018_arXiv}, who noted that the electronic structure of one-dimensional (1D) states at atomic steps in TCIs, revealed by scanning tunneling microscopy \cite{Sessi:2016_S,Iaia:2018_arXiv}, is significantly richer than anticipated previously \cite{Sessi:2016_S,Iaia:2018_arXiv,Polley:2018_ACSN,Rechcinski:2018_PRB}.  According to BWH, these 1D states may show a low-temperature Peierls-like instability leading to the appearance of low-energy excitations associated with topological states at the domains walls of the collective phase. We discuss to what extent our data are consistent with the BWH model and a possible microscopic nature of the collective phase.

{\em Samples.} We investigate here single crystals of rock salt Pb$_{1-y}$Sn$_{y}$Te and Pb$_{1-y-x}$Sn$_{y}$Mn$_{x}$Te obtained via the self-selecting vapor growth method \cite{Dziawa:2012_NM,Szczerbakow:2005_PCGCM}  and the Bridgman technique \cite{Story:1986_PRL}, respectively. Results of electric, magnetic, x-ray, and electron transmission microscopy characterization of the studied samples are presented in the Supplemental Material \cite{SM}. We study samples with $y = 0, 0.2, 0.67, 0.74, 0.8$, and 1, which covers both the topologically trivial and non-trivial range, as the TCI phase occurs for $y \gtrsim 0.30$ \cite{Xu:2012_NC}. According to both angle-resolved photoemission \cite{Tanaka:2012_NP,Xu:2012_NC,Polley:2016_PRB} and magnetotransport investigations \cite{Dybko:2017_PRB},  the surface topological cones coexist with bulk states even for high bulk carrier densities specific to these systems. At the same time, the energetic position of the surface-cone neutrality points with respect to the Fermi level depends on the character of the surface states and the degree of surface oxidation, the question under investigations now \cite{Berchenko:2018_ASS,Chang:2018_AM}. The search for metal precipitates by x-ray diffraction and electron microscopy, also in the vicinity of dislocations, has not revealed the presence of any nanoclustering within the state of the art resolution \cite{SM}.

\begin{figure}[tb]
	\includegraphics[width=9cm]{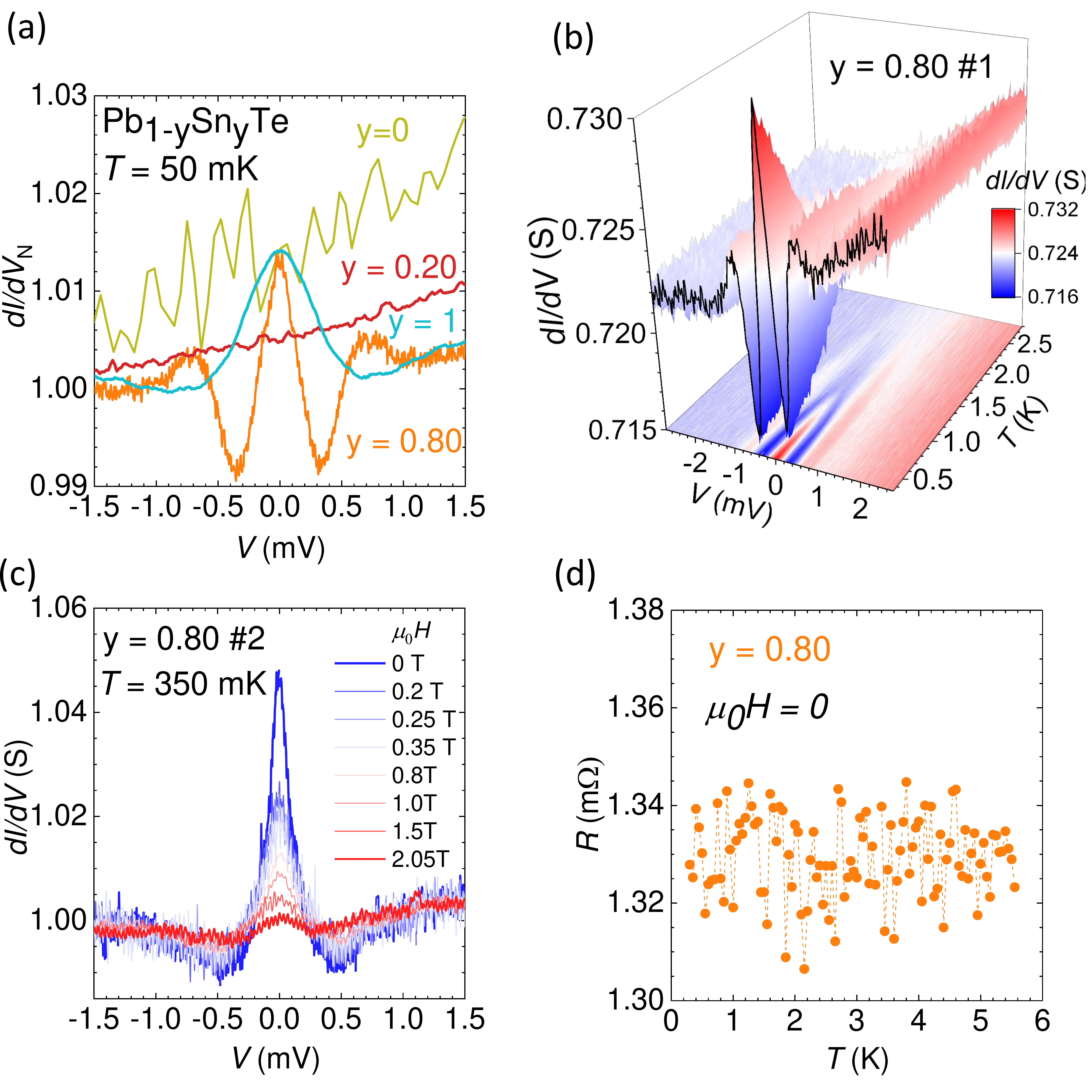}
	\caption{(a) Differential conductance d$I$/d$V$ at 50\,mK normalized to its value at the normal state for the as-grown (001) Pb$_{1-y}$Sn$_y$Te with $y = 0, 0.20, 0.80$, and 1.  The spectrum is featureless for $y = 0$ (PbTe) and $y =0.20$  but shows zero-energy mode characteristics for Sn content ($y = 0.80$ and 1, i.e., SnTe) corresponding to topological crystalline insulator phase. Evolution of the spectrum with temperature (b) and the magnetic field (c) for $y = 0.80$ and two locations of the point contact on the sample surface, respectively. Magnetic field is applied perpendicularly to the (001) plane. (d) Resistance of this sample measured by a four contact method with current density as low as 2.5$\cdot 10^{-3}$\,A/cm$^2$. No global superconductivity is detected.}
\label{fig:ZBCP_1}
\end{figure}

\begin{figure}[tb]
	\includegraphics[width=9cm]{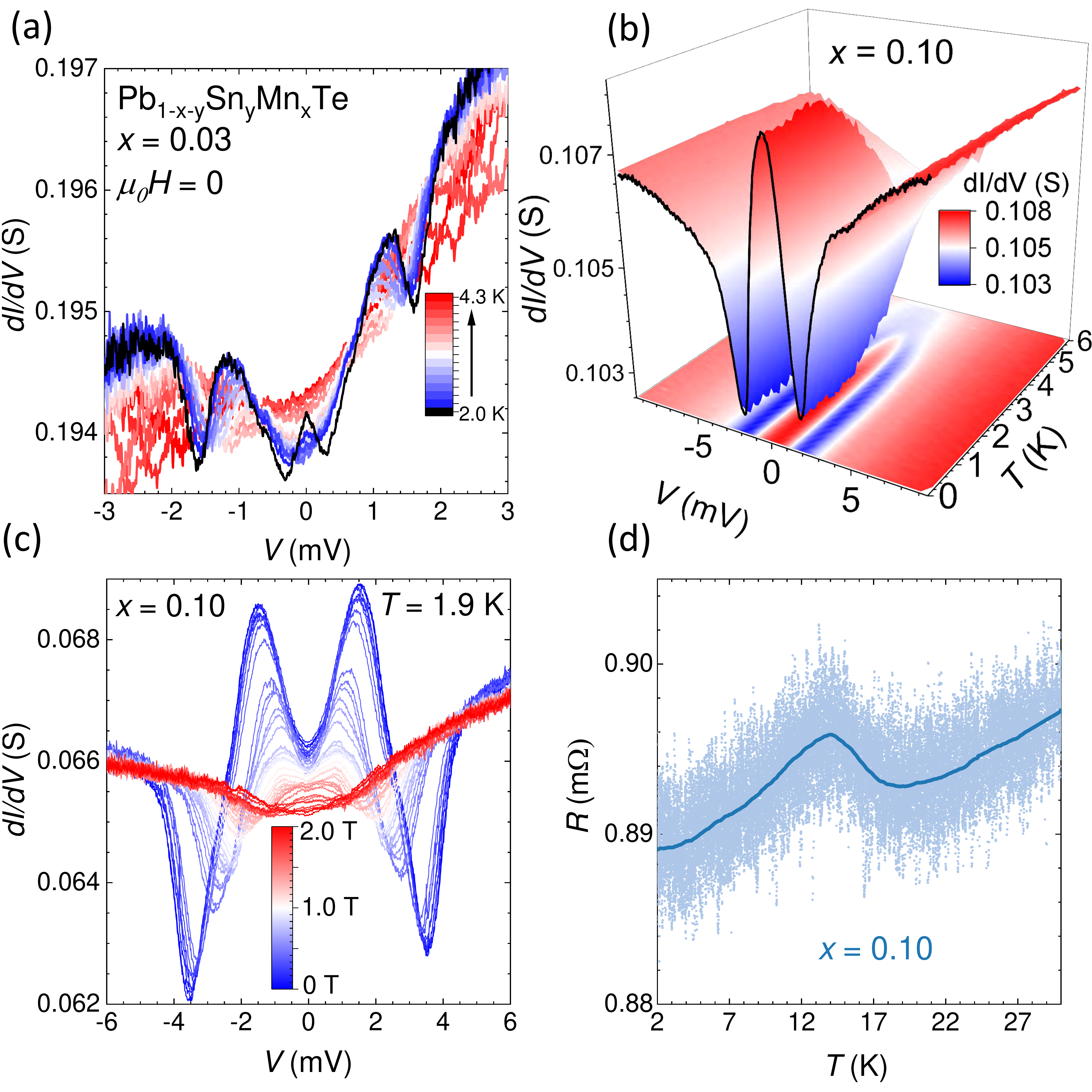}
	\caption{Temperature dependence of differential conductance spectra for the etched (011) Pb$_{0.30}$Sn$_{0.67}$Mn$_{0.03}$Te (a) in a spectroscopic regime and (011) Pb$_{0.16}$Sn$_{0.74}$Mn$_{0.10}$Te (b) in a thermal regime, respectively. (c) Evolution of the spectrum with the magnetic field at 1.8\,K for the cleaved (001) Pb$_{0.16}$Sn$_{0.74}$Mn$_{0.10}$Te in a spectroscopic regime. A magnetic field is applied perpendicularly to the sample plane. (d) Resistance of Pb$_{0.16}$Sn$_{0.74}$Mn$_{0.10}$Te measured by a four contact method with current density $2.5\cdot10^{-5}$\,A/cm$^2$ (noisy trace), and the solid line represents a numerical average over 40 temperature scans. Critical scattering at the Curie temperature $T_{\text{Curie}} =14$\,K is observed but no global superconductivity is detected.}
\label{fig:ZBCP_2}
\end{figure}

According to the results of magnetization measurements \cite{SM}, carried out by employing a superconducting quantum interference device (SQUID), non-magnetic compounds show a field-independent  diamagnetic susceptibility, enhanced by strong interband polarization in the inverted band structure case. The Mn-doped samples contain a Sn concentration corresponding to the TCI phase and a bulk hole density high enough to populate 12 $\Sigma$ valleys. A large density of states (DOS) associated with these valleys makes hole-mediated exchange coupling between Mn ions sufficiently strong to drive the ferromagnetic ordering \cite{Story:1986_PRL,Swagten:1988_PRB}. The Curie temperature $T_{\text{Curie}}$, separating the paramagnetic and ferromagnetic phase, is 2.7 and 14\,K for the Mn concentration $x = 0.03$ and 0.10, and Sn content $y = 0.67$ and 0.74, respectively \cite{SM}. These values are consistent with the mean-field $p$--$d$ Zener model \cite{Dietl:2000_S,Dietl:2014_RMP,SM}.

{\em Point-contact spectroscopy}. We employ the soft point-contact method \cite{Daghero:2010_SST,Sasaki:2011_PRL,Sasaki:2012_PRL}, in which a 15\,$\mu$m Au wire is fixed by silver paint. As shown in Fig.\,\ref{fig:ZBCP_1}(a), d$I$/d$V(V)$ is featureless in the case of the topologically trivial materials PbTe and Pb$_{0.80}$Sn$_{0.20}$Te. However, in the case of the diamagnetic TCI Pb$_{0.20}$Sn$_{0.80}$Te we find at low temperatures $T < T_c$ and magnetic fields $H < H_c$ maxima (ZBCP) centered at $V = 0$ and superimposed on a conductance gap (see also data Ref.\,\onlinecite{SM}).  As observed in other systems \cite{Das:2016_APL, Aggarwal:2016_NM, Wang:2016_NM,Aggarwal:2017_NC,Wang:2017_SB,Wang:2018_SB,Shvetsov:2019_PRB,Zhu:2018_arXiv}, the spectra form vary from contact to contact \cite{SM}. In the thermal regime they change over time in a jump way as well as can be modified by current pulses across the point contact \cite{SM,Daghero:2010_SST}.

\begin{figure*}[tb]
	\includegraphics[width=18cm]{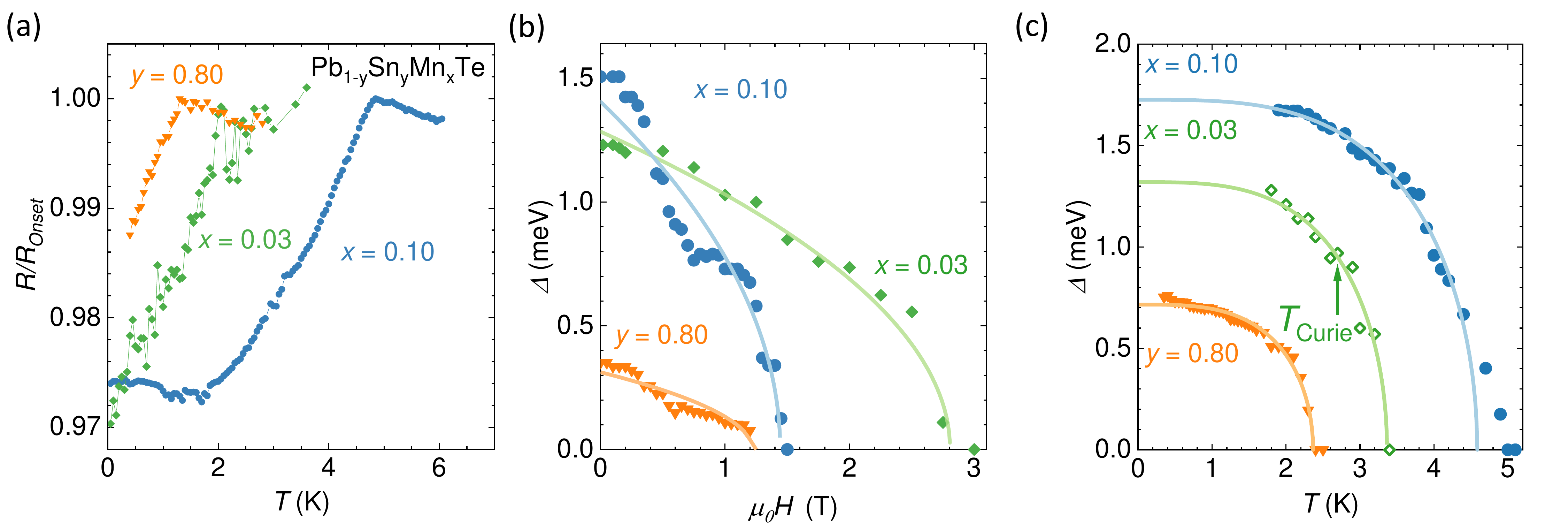}
	\caption{(a) Temperature dependence of the point-contact resistance at the limit of zero bias pointing to a phase transition. (b,c) Conductance gap $\Delta$  evaluated from the differential conductance spectra for samples presented in Figs.\,\ref{fig:ZBCP_1} and \ref{fig:ZBCP_2} for  Pb$_{1-y-x}$Sn$_y$Mn$_x$Te corresponding to the topological crystalline insulator phase vs. magnetic field perpendicular to the surface plane and temperature, respectively. Solid  lines in (b) are fits to $\Delta(T,H) = \Delta(T, H =0)(1- H/H_c)^{1/2}$.  Solid lines in (c) are fits of the BCS formula for $\Delta(T)$ to the experimental points treating $T_c$ and $C$ as adjustable parameters ($C=4.35$, $4.53$, and $3.49$ from top to bottom respectively; $C = 1.76$ in the BCS theory).}
\label{fig:Delta}
\end{figure*}

Within the point-contact theories \cite{Daghero:2010_SST}, the enhanced junction conductance at $T< T_c$ and at $V \approx 0$, i.e., the presence of ZBCP, can be interpreted in terms of the Andreev reflection (pointing to superconductivity) or to enlarged DOS due to the appearance of, for instance,  zero modes at $T< T_c$ \cite{Brzezicki:2018_arXiv}. In either of these scenarios current heating at higher bias voltages may result in side minima  \cite{Sheet:2004_PRB}. Within the superconductivity models, such a spectrum is typical for the thermal regime, i.e., when  the contact diameter is larger than the inelastic diffusion length \cite{Daghero:2010_SST} [Fig.~\ref{fig:ZBCP_1}(c)]. In the opposite limit, spectroscopic information is not blurred, and side maxima in d$I$/d$V$, reflecting the DOS enlargement at the gap edges,  provide the value of the relevant gap. Within such an approach the data in Fig.~\ref{fig:ZBCP_1}(b) (see also the data in Ref.\,\cite{SM}), correspond to the spectroscopic regime. An interesting question arises on whether such a phenomenology can be directly applied to other gaped collective states, such as the one proposed by BWH \cite{Brzezicki:2018_arXiv}.

According to Fig.\,\ref{fig:ZBCP_2},  d$I$/d$V$ shows clear spectroscopic features in magnetic crystals. The spectrum for Pb$_{0.3}$Sn$_{0.67}$Mn$_{0.03}$Te [Fig.\,\ref{fig:ZBCP_2}(a)] exhibits a gap with a small zero-bias peak vanishing smoothly with increasing temperature. For a cleaved (100) surface of Pb$_{0.16}$Sn$_{0.74}$Mn$_{0.10}$Te, a splitting of ZBCP, resembling Andreev reflection characteristics in the spectroscopic regime and for a non-zero barrier transparency \cite{Daghero:2010_SST} has been detected [Fig.\,\ref{fig:ZBCP_2}(c)].

Importantly, temperature and magnetic field ranges in which these spectroscopic features appear are rather enhanced compared to non-magnetic crystals.  This is best seen in Fig.\,\ref{fig:Delta} that depicts the contact resistance and a half of the energy distance between conductance maxima, $\Delta(T,H)$ for all studied samples with high Sn content.  Furthermore, according to Fig.\,\ref{fig:Delta}, $\Delta(T)$ is quite well described by an interpolation formula of the BCS expression, $\Delta(T)= C k_{\text{B}} T_c[1-(T/T_c)^{3.3}]^{0.5}$, where $C$ in our case is more than twofold greater than the BCS value $C = 1.76$. Similarly, a reasonable account of $\Delta(H)$ data is obtained by using another interpolation formula suitable for type II superconductors,  $\Delta(T,H) = \Delta(T, H = 0)(1 - H/H_c)^{1/2}$. This description of $\Delta(T,H)$ holds for the diamagnetic  Pb$_{0.2}$Sn$_{0.80}$Te, ferromagnetic Pb$_{0.16}$Sn$_{0.74}$Mn$_{0.10}$Te, and across the paramagnetic--ferromagnetic  phase boundary, the case of Pb$_{0.20}$Sn$_{0.67}$Mn$_{0.03}$Te, in which $T_c > T_{\text{Curie}}$. The existence of a transition to another phase is also documented by a kink in the temperature dependence of the differential resistance for $V\rightarrow 0$ at $T_c$, as shown in Fig.\,\ref{fig:Delta}(a).

The accumulated results demonstrate, therefore, the appearance in junctions of the normal metal with diamagnetic, paramagnetic, and ferromagnetic IV-VI TCIs of another phase below $T_c$ and $H_c$, which is characterized by an energy gap and excitations residing near its center.  The presence of a phase transition rules out the Kondo effect \cite{Appelbaum:1966_PRL,Anderson:1966_PRL,Meir:1993_PRL,Pustilnik:2001_PRL} and the Coulomb gap \cite{Efros:1975_JPC,Altshuler:1979_JETP} as the mechanisms accounting for the observed features.  Similarly, the absence of the global superconductivity without metal layers [Figs.\,\ref{fig:ZBCP_1}(d) and \ref{fig:ZBCP_2}(d) as well as data in Ref.\,\onlinecite{SM}] indicate that unconventional 2D superconductivity associated with surface topological states \cite{Kundu:2017_PRB} or dislocation arrays \cite{Tang:2014_NP} in TCIs does not appear under our experimental conditions. The absence of 2D superconductivity is also documented by similar values of $H_c$ for the magnetic field perpendicular and parallel to the surface \cite{SM}.

\begin{figure}[t]
	\includegraphics[width=9cm]{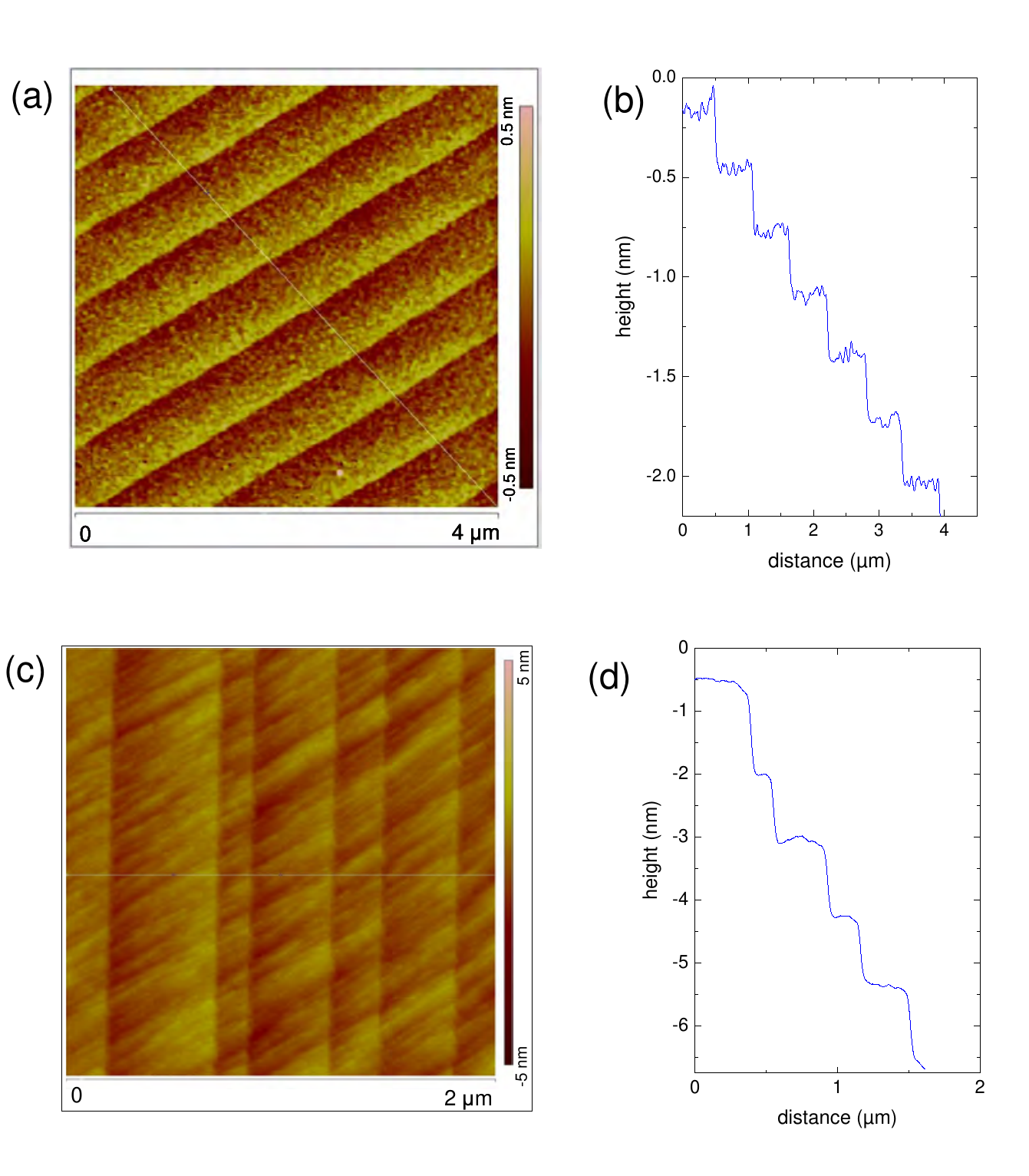}
	\caption{AFM images of the studied single crystal surfaces. (a) Naturally grown (001) facet of Pb$_{0.20}$Sn$_{0.80}$Te showing surface steps. (c) Cleaved (001) surface of Pb$_{0.16}$Sn$_{0.74}$Mn$_{0.10}$Te showing multilayer steps and higher roughness.  (b) and (d) depict step height profiles. The obtained values of about 0.3\,nm in (b) correspond to a single atomic step (315\,pm).}
	\label{fig:AFM}
\end{figure}

{\em Role of superconducting precipitates.}
According to comprehensive structural and SQUID studies, magnetic nanoparticles account for high $T_{\text{Curie}}$ ferromagnetism observed in a number of semiconductors and oxides over the last two decades \cite{Dietl:2015_RMP}. Similarly to the ferromagnetic case, embedded nanoparticles can show a variety of superconducting characteristics, such as the Meissner effect, which may depend on nanoparticle chemical composition, strain, and size \cite{Darchuk:1998_Semicon,Murase:1986_SS,Fogel:2006_PRB,Li:2003_PRB,Yeh:2008_CSA}. Here, the presence of superconducting precipitates that could give rise to local superconductivity is ruled out, within an experimental accuracy of 0.1\,ppm,  by high-sensitivity beyond state-of-the-art SQUID magnetometry \cite{SM,Gas:2019_MST}. Furthermore, no precipitates have been revealed by state-of-the-art x-ray diffraction and transmission microscopy measurements \cite{SM}. Similarly, no indications of superconductivity have been found in topological samples covered entirely by silver paint or containing a deposited silver film \cite{SM}, pointing to the absence of both interfacial superconductivity and superconducting inclusions at the metal/semiconductor interface.

{\em Role of surface atomic steps.}
Figure \ref{fig:AFM} presents the surface morphology of our single crystals under ambient conditions, determined by atomic force microscopy (AFM). As seen in Figs.\,\ref{fig:AFM}(a,b), (001) facets of as grown Pb$_{0.20}$Sn$_{0.80}$Te contain atomically flat 0.5\,$\mu$m wide terraces, terminated by monoatomic steps. Similarly, Figs.\,\ref{fig:AFM}(c,d) visualize a (001) surface of cleaved Pb$_{0.16}$Sn$_{0.74}$Mn$_{0.10}$Te showing larger  roughness and multilayer steps. Results in Figs.\,\ref{fig:ZBCP_1}(b,c) and \ref{fig:ZBCP_2}(c) were actually taken for these two surfaces, respectively. We claim that a collective low-temperature phase of the carrier liquid occupying 1D topological step states \cite{Sessi:2016_S,Brzezicki:2018_arXiv,Polley:2018_ACSN,Rechcinski:2018_PRB,Iaia:2018_arXiv} may account for the differential conductance spectra reported here. According to AFM  micrographs, a single Ag grain extends over a dozen of steps.  The magnitude of ZBCP, about 50 to $500e^2/h$, is consistent with the fact that several grains participate in the charge transport process.

The above interpretation requires the Fermi level of our $p$-type samples to reside within the 1D states, whereas tight-binding computations place the 1D band in the gap \cite{Sessi:2016_S,Brzezicki:2018_arXiv,Polley:2018_ACSN,Rechcinski:2018_PRB,Iaia:2018_arXiv}. We note, however, that our experimental method implies the formation of Schottky's metal-semiconductor junction, which leads usually to the Fermi level pinning in the band gap region at the semiconductor surface \cite{Tung:2014_APR}. This effect, together with a considerable width of the 1D band, as predicted by the BWH theory \cite{Brzezicki:2018_arXiv}, make the partial occupation of the 1D states plausible. The depletion in the bulk carrier density implies also that Mn spins adjacent to the surface will be rather coupled by antiferromagnetic superexchange \cite{Gorska:1988_PRB} than by the hole-mediated ferromagnetic interactions dominating in the bulk \cite{Story:1986_PRL,SM}.

However, since the Fermi liquid is unstable in the 1D case, one expects the emergence of a collective state at low temperatures driven by carrier correlation, presumably enhanced by coupling to phonons and localized spins.  For a class of electronic instabilities, such as superconductivity and charge/spin density waves, the BCS relation between the gap $2\Delta$ and the critical temperature $T_c$ remains valid within the mean-field approximation. However, because of the dominating role played by thermal and quantum fluctuations of the order parameter, the apparent magnitude of $T_c$ becomes much reduced in the 1D case \cite{Arutyunov:2008_PR,Pouget:2016_CRP}. This explains the enhanced magnitude of $C$ over the BCS value (Fig.\,\ref{fig:Delta}) but indicates also that 1D surface states may not support superconductivity with $T_c$ as high as 5\,K.

As demonstrated  by BWH \cite{Brzezicki:2018_arXiv}, the 1D states adjacent to the surface atomic step show much more abundant spectrum than anticipated previously \cite{Sessi:2016_S,Polley:2018_ACSN,Rechcinski:2018_PRB,Iaia:2018_arXiv}. Within this insight, the effects of magnetic instabilities have been considered and the nature of low-energy excitations proposed \cite{Brzezicki:2018_arXiv}. In an analogy with the Su--Schrieffer--Heeger model \cite{Heeger:1988_RMP}, the  low-energy modes are associated with domain walls between the regions characterized by the opposite directions of the order parameter for which, as proven, topological invariants differ. These walls, and thus, low energy excitations vanish in the magnetic field with a rate determined by a competition of the carrier-carrier exchange coupling with the Zeeman, spin-orbit, and $sp$--$d$ exchange interactions. Depending on the assumed broadening, the evaluated conductance spectra show a single ZBCP, or a more complex peak structure that reflects the multi-mode excitation spectrum and may resemble Andreev reflection \cite{Brzezicki:2018_arXiv}.

{\em Summary and outlook.} Our results on soft point-contact spectroscopy bring into light the existence of a low-temperature phase in lead-tin and lead-tin-manganese tellurides with tin content corresponding to the topological phase, which substantiates the universality of the phenomenon \cite{Das:2016_APL,Aggarwal:2016_NM,Wang:2016_NM,Aggarwal:2017_NC,Wang:2017_SB,Wang:2018_SB,Shvetsov:2019_PRB,Zhu:2018_arXiv}. The findings could be explained by the presence of superconducting nanoparticles, the surface topological states playing only an ancillary role.
However, no indications for the formation of nanoparticles have so far been found. At the same time, our experiments do not provide evidence for the presence of superconductivity at the interface between the metal and the TCI. Hence, this phase can be also linked to carries occupying 1D topological states adjacent to surface atomic steps, which undergo a transition to a collective state at sufficiently low temperatures and magnetic fields. Four microprobe measurements of conductance along individual steps in materials with the Fermi level within the step states are expected to demonstrate whether this collective phase is a 1D superconductor or a 1D gaped insulator with low-energy excitations at the domain walls. Future work will show to what extent the unusual properties associated with the presence of 1D and 0D topological states offer new and hitherto unexplored functionalities.


{\em Acknowledgements.}
We thank Marek Foltyn and Pawel Skupinski for technical support and Victor Galitski for a valuable discussion. The Research Foundation MagTop -- International Centre for interfacing Magnetism and Superconductivity with Topological Matter (short name: International Research Centre MagTop) is funded by the Foundation for Polish Science through the IRA Programme financed by theEU within SG OP Programme. The work at the Institute of Physics, Polish Academy of Sciences was supported by the National Science Center (Poland) through the following grants: PRELUDIUM (2015/19/N/ST3/02626), OPUS (2012/07/B/ST3/03607, 2013/09/B/ST3/04175,  2014/15/B/ST3/03833, 2017/27/B/ST3/02470), and MAESTRO (2011/02/A/ST3/00125). The TEM/FIB investigation was performed on equipment supported by Polish  Government under Agreement 4277/E-67/SPUB/2017/1.

\clearpage

{\section*{Supplemental Information}}

\tableofcontents
\renewcommand{\thefigure}{S\arabic{figure}}
\setcounter{figure}{0}
\renewcommand{\thesubsection}{S\arabic{subsection}}
\setcounter{subsection}{0}

%
\subsection{Studied samples}

Table I presents information on the studied samples.

\begin{table}[h]
		\begin{tabular}{c|cc}
			Sample & $p$\,(cm$^{-3}$) & $\mu$\,(cm$^{2}$/Vs) \\
			\hline
			PbTe & $5\times 10^{18}$ & 970 \\
			Pb$_{0.80}$Sn$_{0.20}$Te & $4.9\times 10^{19}$ & 280 \\
			Pb$_{0.20}$Sn$_{0.80}$Te & $7.5\times 10^{20}$ & 80 \\
			SnTe & $3\times 10^{20}$ & 350\\
			Pb$_{0.30}$Sn$_{0.67}$Mn$_{0.03}$Te & $2.6\times 10^{20}$ &  110 \\
			Pb$_{0.16}$Sn$_{0.74}$Mn$_{0.10}$Te & $1.4\times 10^{21}$ & 180 \\
		\end{tabular}
		\caption{Chemical composition, hole concentration and mobility from
			room temperature Hall measurements of the investigated samples.}
\end{table}

\subsection{Search for precipitates by x-ray diffraction}
\begin{figure}[bt]
	\includegraphics[width=6cm]{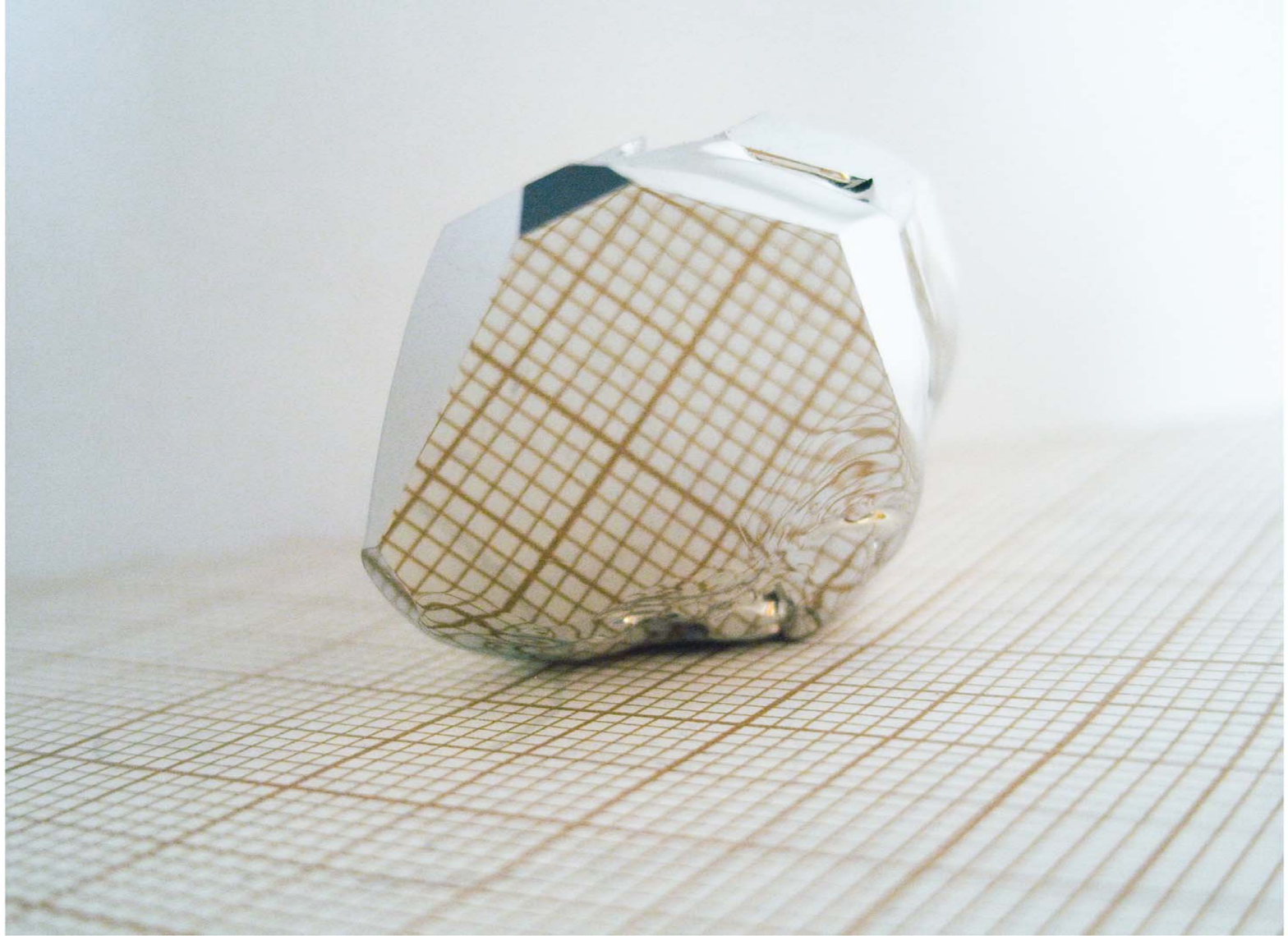}
	\caption{A photograph of an as-grown SnTe crystal with (100) and (111) facets.}
	\label{fig:SnTe_photo}
\end{figure}

For our experiments we prepared dedicated bulk crystals of the highest, currently available, quality. Our SnTe and Pb$_{0.2}$Sn$_{0.8}$Te crystals were grown by self-selecting vapor growth (SSVG) method known to provide large (volume of the order of 1 cm$^3$) single crystals with dominant (001) and additional (111) crystal facets (see photo in Fig.~\ref{fig:SnTe_photo}). The SSVG growth involves vapor transport in the presence of a small temperature gradient at temperatures slightly below the melting point, and provides IV-VI semiconductor crystals of excellent crystal quality, good chemical homogeneity with a relatively small deviation from stoichiometry, and free from detectable inclusions of foreign crystal phases.

Due to low vapor pressure of Mn,  the growth of magnetic Pb$_{0.16}$Sn$_{0.74}$Mn$_{0.10}$Te crystals requires the application of the Bridgman method, i.e., crystallization from a melt. Although crystalline grains are here much smaller (typical dimension 0.5 mm) they are only slightly misoriented (0.06$^{\rm o}$). The uniformity of both carrier density and ferromagnetic properties point to the homogeneity of the chemical composition.

\begin{figure}[bt]
	\includegraphics[width=8cm]{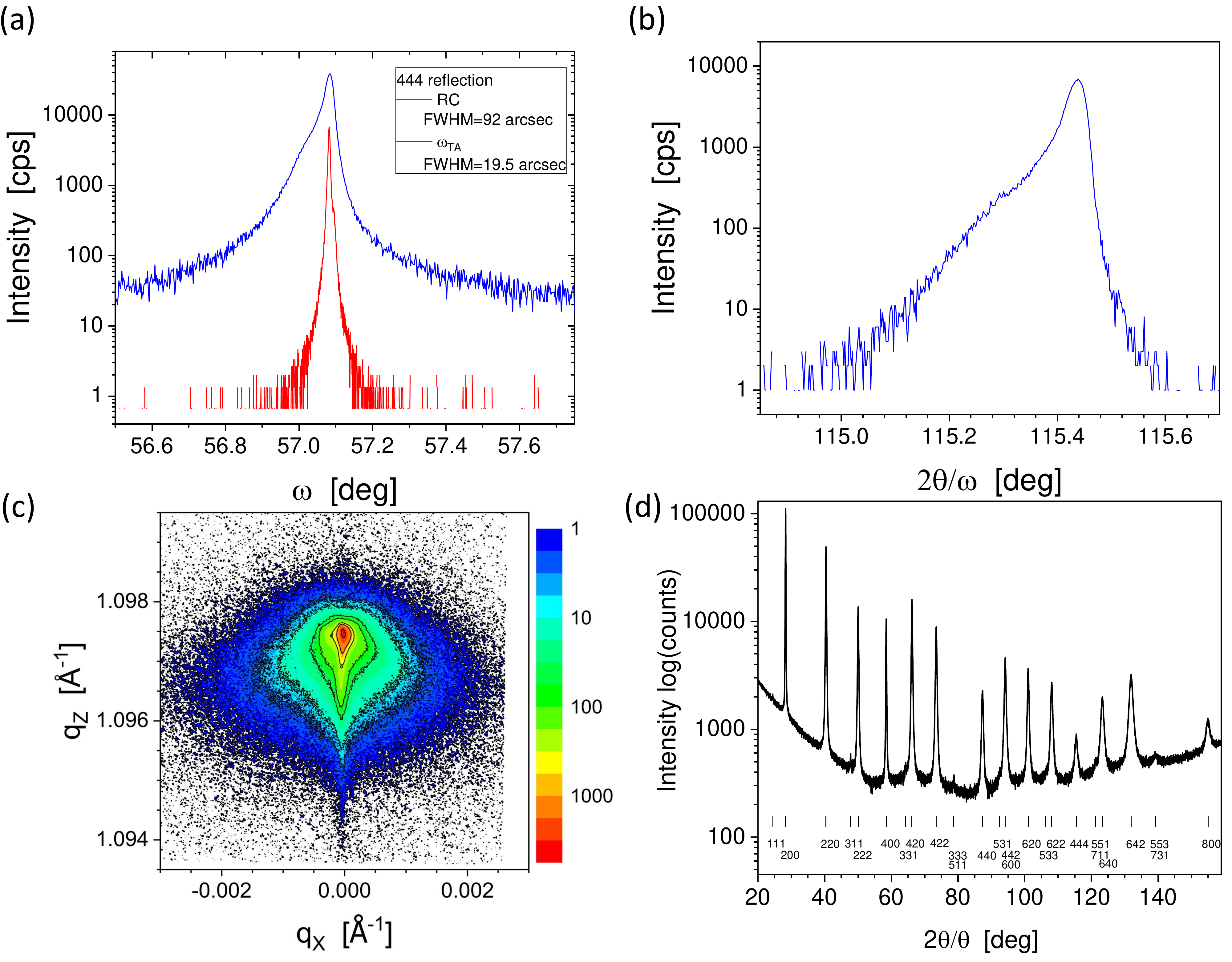}
	\caption{(a-c) High resolution x-ray diffraction of SnTe; reflection 444; sample reflected x--ray beam size  $1\times1$\,mm$^2$: (a) $\omega$-scan measurements; upper line - rocking curve, lower - $\omega_{\rm TA}$-scan; (b) 2$\theta$/$\omega$ scan; (c) reciprocal space map. The logarithmic scale is used, intensity is shown in counts per second [cps]. (d) Powder diffraction patterns of SnTe. The vertical bars indicate positions of the Bragg peaks.}
	\label{fig:xraySnTe}
\end{figure}

\begin{figure}[bt]
	\includegraphics[width=8cm]{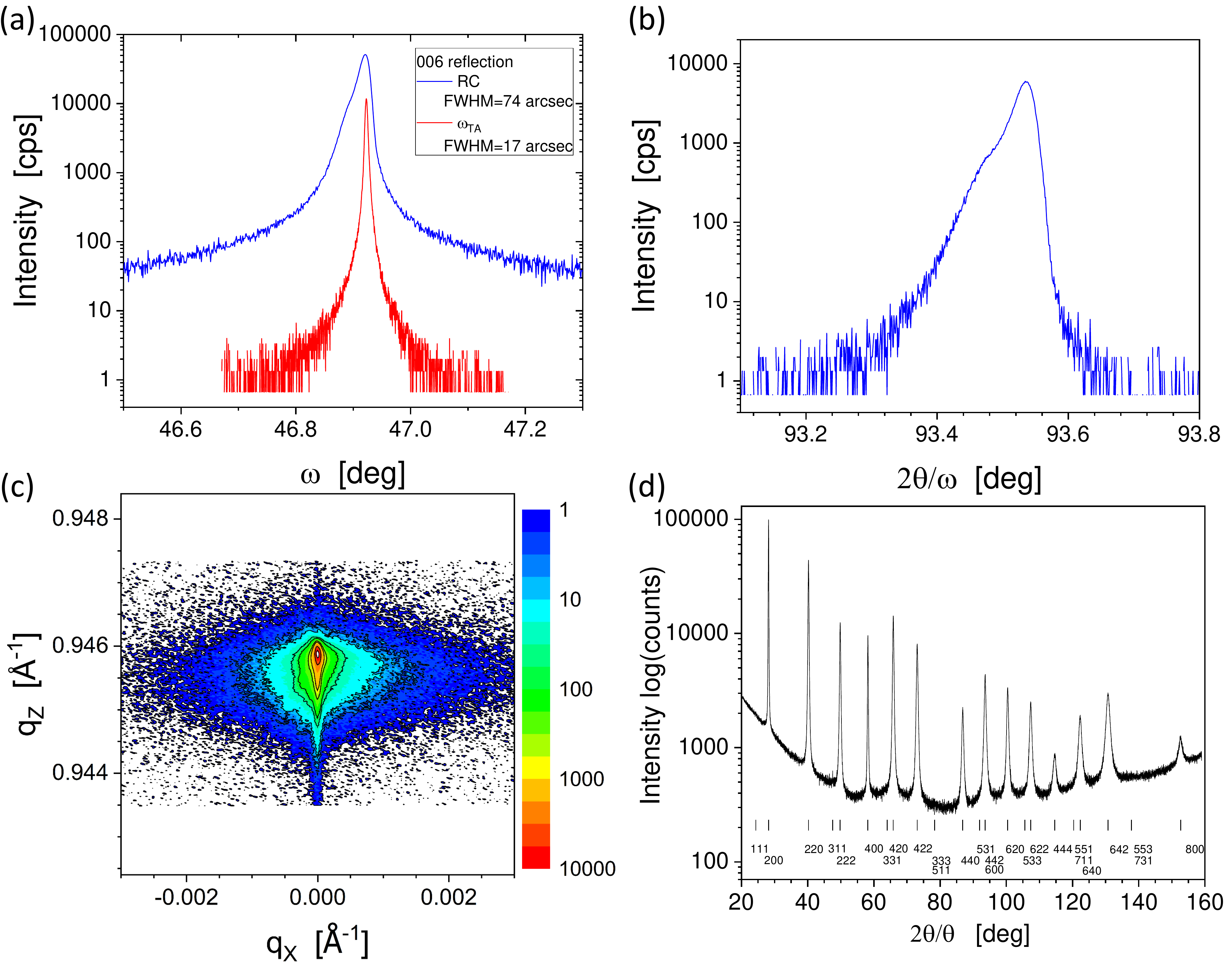}
	\caption{The same as in Fig.~S2 but for Pb$_{0.2}$Sn$_{0.8}$Te; reflection 006; sample reflected x--ray beam size  $1\times1$\,mm$^2$.}
	\label{fig:xrayPbSnTe}
\end{figure}

\begin{figure}[bt]
	\includegraphics[width=8cm]{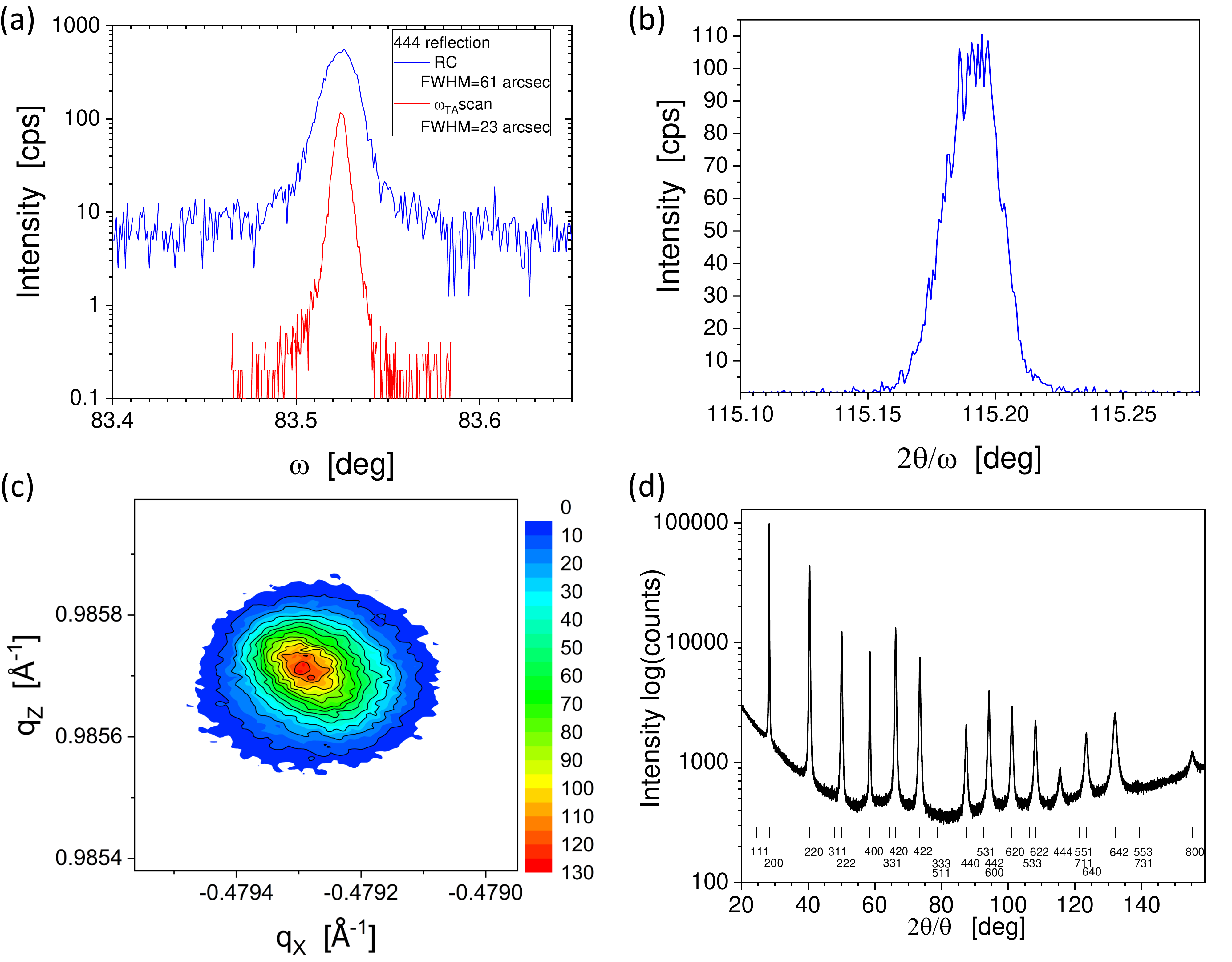}
	\caption{The same as in Fig.~S2 but for Pb$_{0.16}$Sn$_{0.74}$Mn$_{0.10}$Te; reflection 444; sample reflected x--ray beam size  $0.5\times1$\,mm$^2$.}
	\label{fig:xrayPbSnMnTe}
\end{figure}

The crystallographic quality of the samples has been assessed by the-state-of-the-art high-resolution x--ray diffraction (HR-XRD) and powder diffraction. The $\omega$, 2$\theta$/$\omega$ scans, and reciprocal space maps  (RSM) have been collected for reflection 444 (SnTe and Pb$_{0.16}$Sn$_{0.74}$Mn$_{0.10}$Te) and 006 (Pb$_{0.20}$Sn$_{0.80}$Te), as shown in Figs.~\ref{fig:xraySnTe},\,\ref{fig:xrayPbSnTe} and \ref{fig:xrayPbSnMnTe}, respectively. For the crystals of SnTe and Pb$_{0.2}$Sn$_{0.8}$Te grown by the SSVG method one finds very sharp diffraction peaks fully accounted for by a single-rock salt crystal phase. The application of various XRD geometries (as indicated in Figs.~\ref{fig:xraySnTe} and \ref{fig:xrayPbSnTe}) permitted the determination of the RC width parameters of 92 and 74 arcsec for SnTe and Pb$_{0.2}$Sn$_{0.8}$Te, respectively. After correcting for diffraction peaks asymmetry and background contribution due to diffusive scattering by native defects (vacancies) present in the crystals we could also evaluate the crystal perfection of (111) and (001) planes as given by the RC width parameter 17-19.5 arcsec. One may notice that this magnitude of the RC parameters are observed for good GaAs (001) crystals.

In the case of Pb$_{0.16}$Sn$_{0.74}$Mn$_{0.10}$Te - the planes (111) were inclined at the angle of 25$^{\rm o}$ to the surface, so that 444 was asymmetrical (the x--ray beam geometry at a high angle to the sample was used).  The Pb$_{0.16}$Sn$_{0.74}$Mn$_{0.10}$Te sample consists of slightly misoriented $0.5 \times 0.1$\,mm$^2$ single crystalline grains. The mean grain deviation angle is estimated to be $0.06 \pm 0.02^{\rm o}$. For an x--ray beam limited by a mask and slit of such dimensions, the single crystal diffraction pattern is observed (Fig.~\ref{fig:xrayPbSnMnTe}).

\begin{center}
	\begin{table}
		\begin{tabular}{c|c}
			Sample &  a\,($\AA$) \\
			\hline
			Pb$_{0.20}$Sn$_{0.80}$Te &  6.3426(3) \\
			SnTe &  6.3129(3) \\
			Pb$_{0.16}$Sn$_{0.74}$Mn$_{0.10}$Te & 6.3098(5) \\
		\end{tabular}
		\caption{Lattice parameters of investigated samples.}
	\end{table}
\end{center}

Wide-angle x-ray power diffraction for used to detect foreign crystallographic phases in the studied samples. The data collected in Figs.~\ref{fig:xraySnTe}(d), \ref{fig:xrayPbSnTe}(d), and \ref{fig:xrayPbSnMnTe}(d)  were taken with excellent statistics, peak-to-background ratio ~74, and in a broad angular range ($2\theta$ up to 159$^{\rm o}$). The positions of all peaks point to the rock salt crystal symmetry (Fm3-m) with no indication for the presence of a secondary phase. The determined lattice parameters are given in the Table~II.

\begin{figure}[b]
	\includegraphics[width=6cm]{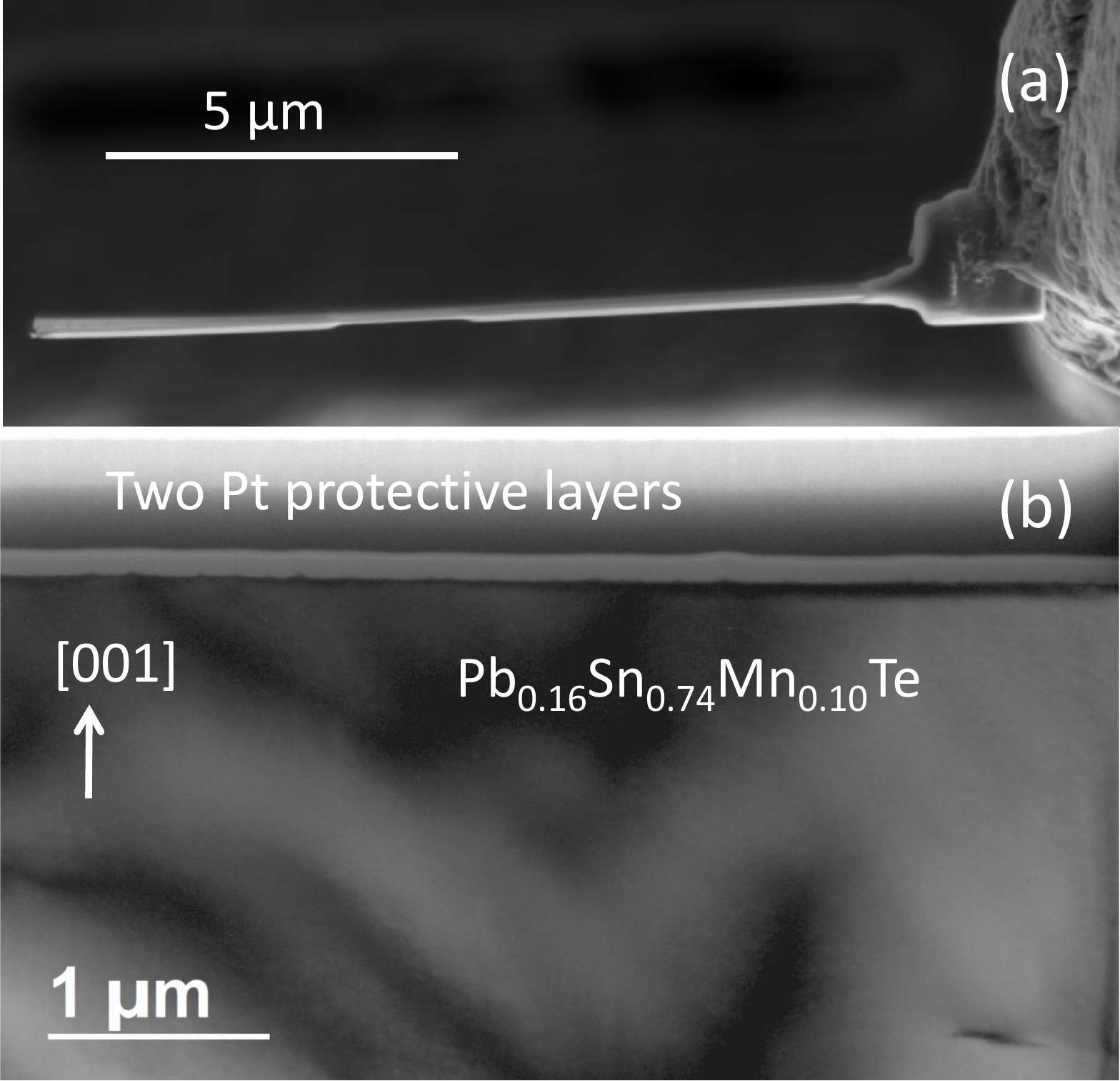}
	\caption{(a) Scanning electron microscope image of the edge of the lamella  attached to the transmission electron microscope (TEM) copper support; (b) TEM bright field image  of a defect-free $5\times2.5$ \,$\mu$m  area of the specimen. }
\label{fig:TEM_1}
\end{figure}

\begin{figure}[tb]
	\includegraphics[width=6cm]{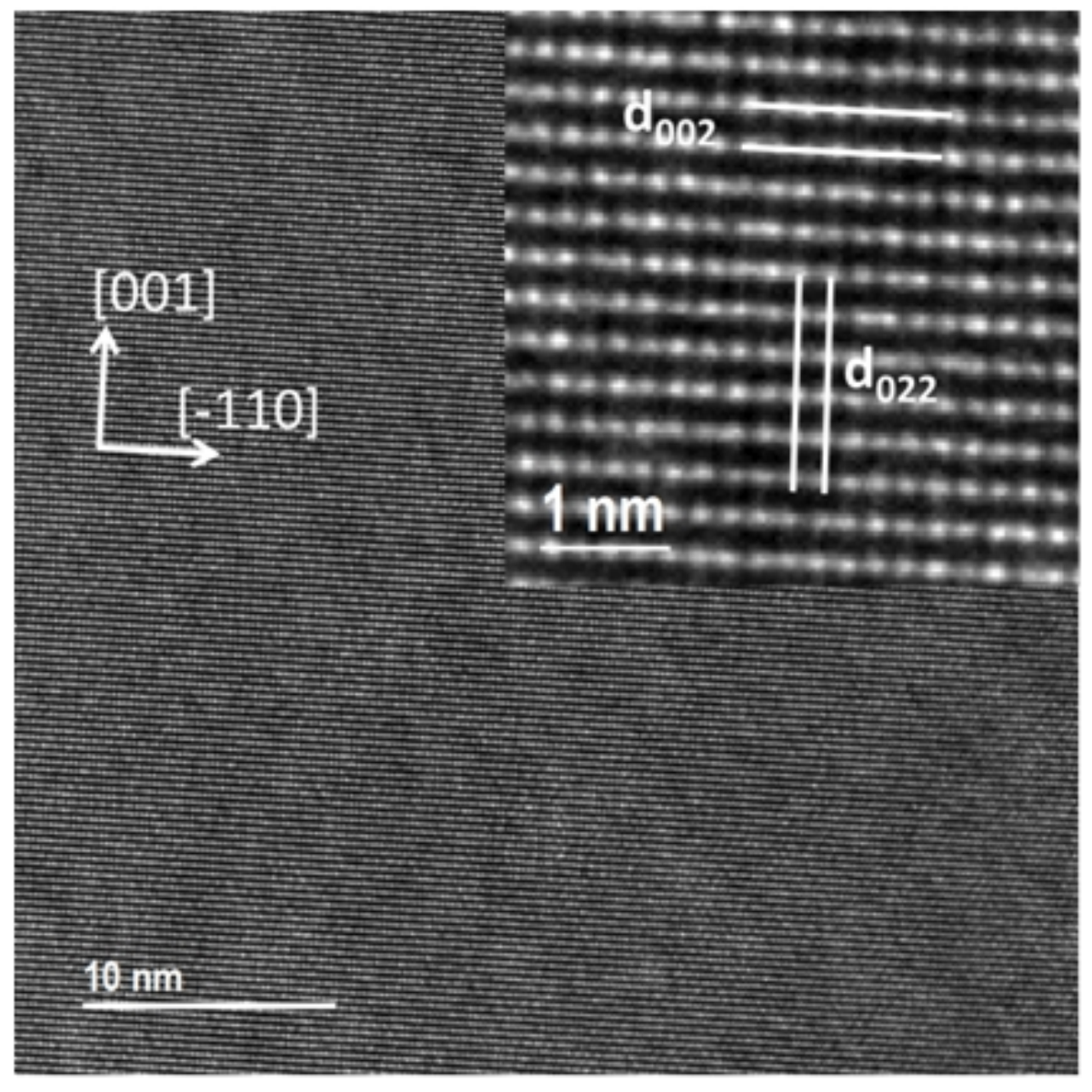}
	\caption{High-resolution transmission electron microscope image of zone axis (110) of Pb$_{0.16}$Sn$_{0.74}$Mn$_{0.10}$Te crystal.}
\label{fig:TEM_2}
\end{figure}

\subsection{Search for precipitates by electron microscopy}

\begin{figure}[tb]
	\includegraphics[width=6cm]{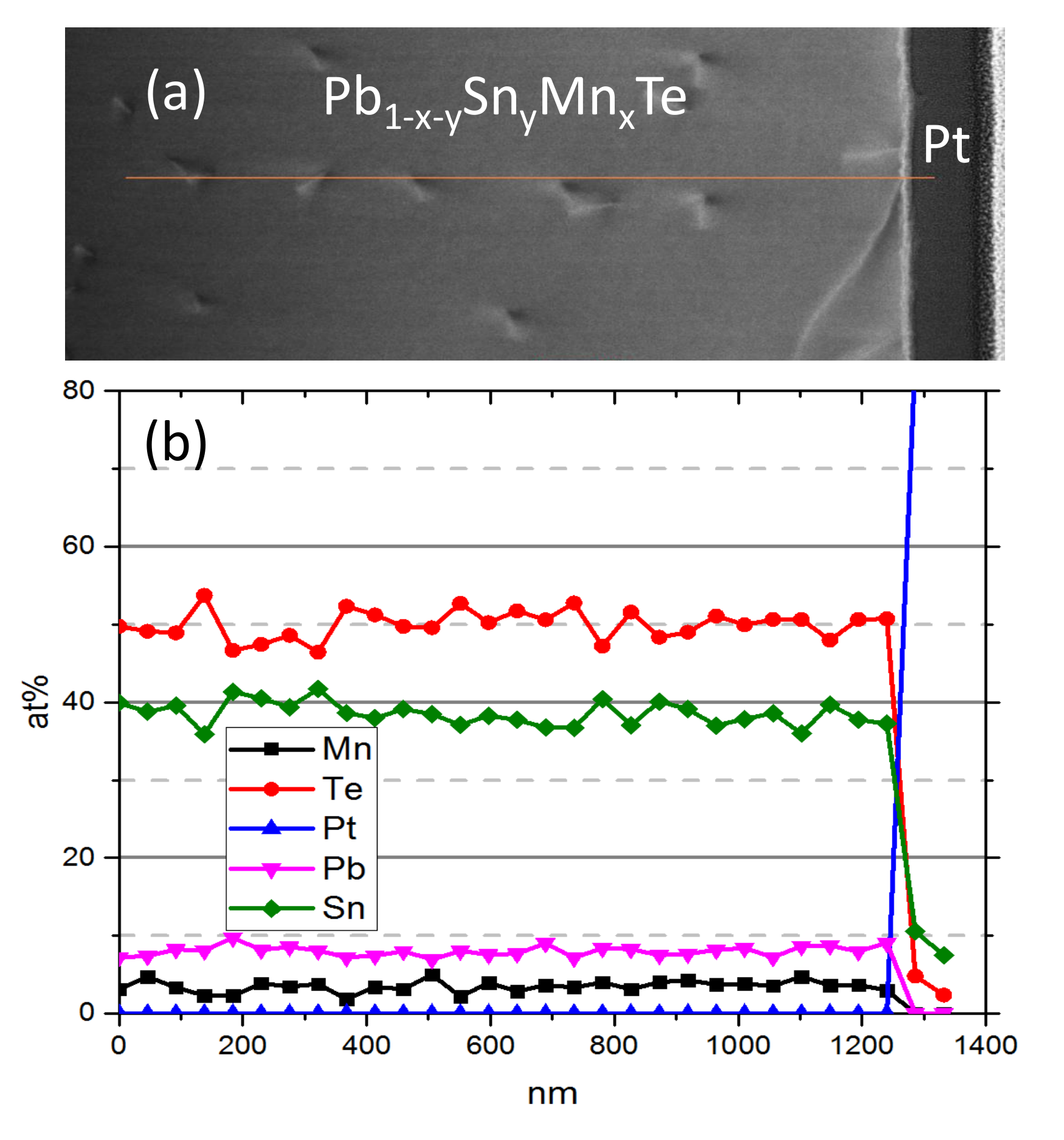}
	\caption{(a) STEM-HAADF (out of zone axis). (b) EDS elemental concentration profile of Sn, Pb, Te, Mn, Pt of the area with near-surface dislocations. Note that atomic, not cation, concentrations are shown.}
\label{fig:TEM_3}
\end{figure}

Figure \ref{fig:TEM_1}(a) shows scanning electron microscopy (SEM) image of a lamella prepared of Pb$_{0.16}$Sn$_{0.74}$Mn$_{0:10}$Te for transmission electron microscopy (TEM) studies. A $5\times2.5$\,$\mu$m defect-free area is presented in Fig.\,\ref{fig:TEM_1}(b). The dark wavy features are due to lamella bending.  The sharp feature seen at the bottom right corner is caused by mechanical deformation that occurred during the thinning process or by processing of the crystal during previous investigations.  Structural defects (mainly small dislocation loops) can be found is some regions near the surface of the crystal - no deeper than 2\,$\mu$m from the surface. No fluctuations of elemental composition or precipitations of other phases are detected. It appears that Pb and Mn atoms substitute the Sn atoms without ordering or segregation that could be detected within our sensitivity. The high-resolution TEM (HRTEM) image (Fig.\,\ref{fig:TEM_2}) indicates a perfect cubic lattice visible in [110] zone axis. A quantitative  profile of the crystal region containing structural defects (dislocations) near the surface zone obtained by energy dispersive x-ray spectroscopy (EDS) is presented in Fig.\,\ref{fig:TEM_3}. No aggregation of particular constituents is detected, as composition fluctuations are within the error bar of the method. The determined atomic concentrations of the constituting elements are in good agreement with results of our SQUID measurements and specifications found in the previous characterization study$^{S7}$.

\begin{figure}[tb]
	\includegraphics[width=6cm]{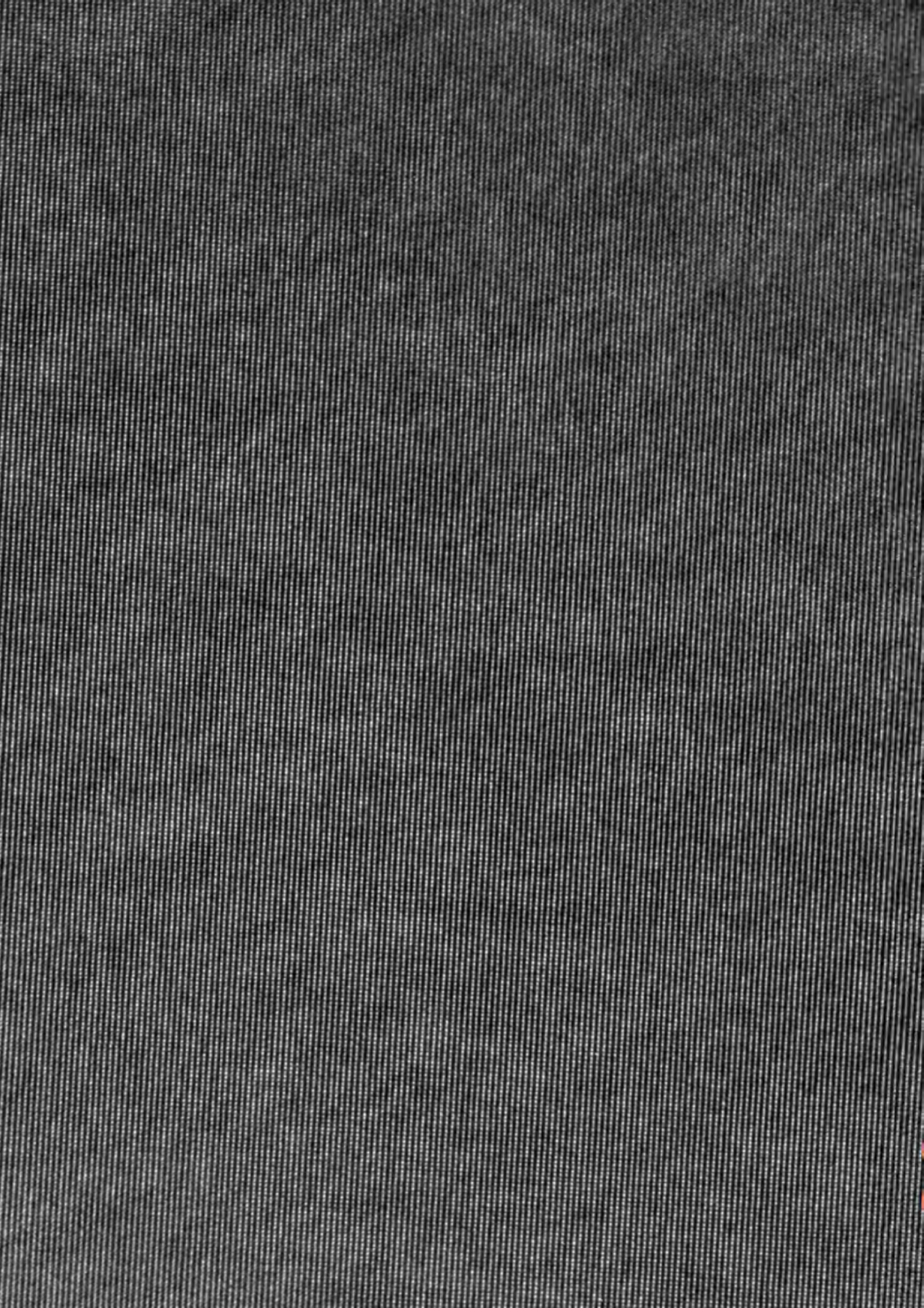}
	\caption{TEM scan of large 71x71 nm of Pb$_{0.16}$Sn$_{0.74}$Mn$_{0.10}$Te crystal. No signs of precipitates or nanoclustering was found in TEM studies.}
	\label{fig:TEM_4}
\end{figure}
\subsection{Magnetic properties}

\begin{figure}[tb]
	\includegraphics[width=8cm]{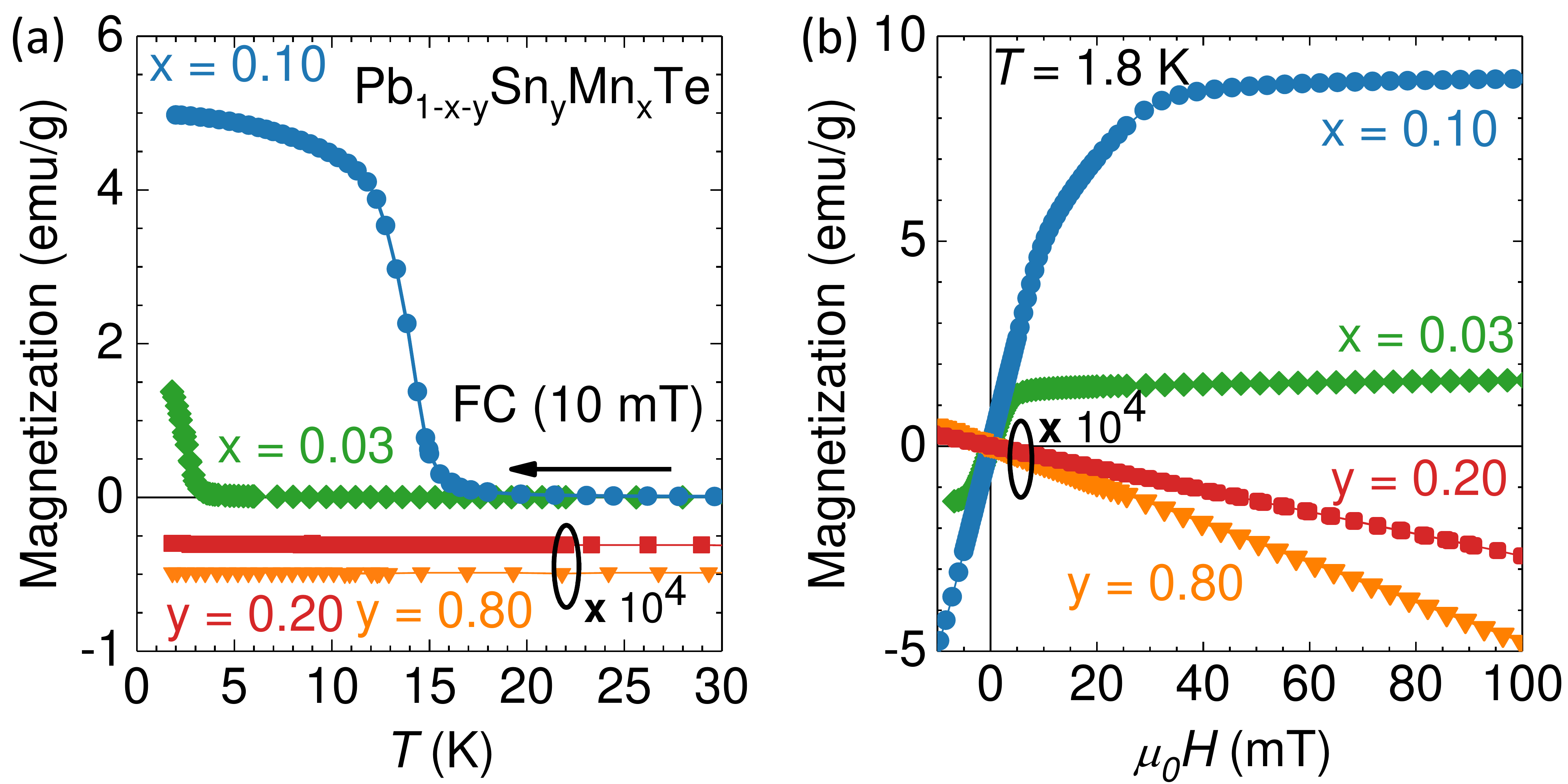}
	\caption{Magnetization as a function of temperature (a) and the magnetic field (b) of Pb$_{1-y-x}$Sn$_y$Mn$_x$Te with various Mn concentrations $x$ and the Sn content $y$. The data show the presence of ferromagnetism at low temperatures in Mn-doped samples with the Curie temperature increasing with $x$. Without Mn doping the samples are diamagnetic -- in order to visualize the diamagnetism magnitude, the magnetization values are multiplied by a factor of $10^4$.}
\label{fig:ferro}
\end{figure}

According to results of magnetization measurements collected in Fig.\,\ref{fig:ferro}, non-magnetic compounds show field-independent  diamagnetic susceptibility, enhanced by strong interband polarization in the inverted band structure case.
The Mn-doped samples contain Sn concentration corresponding to the TCI phase and the hole density high enough to populate twelve $\Sigma$ valleys. A large density of states (DOS) associated with these valleys makes hole-mediated exchange coupling between Mn ions sufficiently strong to drive the ferromagnetic ordering \cite{Story:1986_PRL,Swagten:1988_PRB}. As shown in Fig.\,\ref{fig:ferro}, the Curie temperature $T_{\text{Curie}}$, separating the paramagnetic and ferromagnetic phase, is 2.7 and 14\,K for the Mn concentration $x = 0.03$ and 0.10, and the Sn content $y = 0.67$ and 0.74, respectively.

According to the mean-field $p-d$ Zener model \cite{Dietl:2000_S,Dietl:2014_RMP}
\begin{equation}
T_{\text{Curie}} = x_{\text{eff}}N_0S(S+1)\beta_{\text{eff}}^2\rho_{\text{F}}/12k_{\text{B}} - T_{\text{AF}},
\end{equation}
where $x_{\text{eff}}< x$ and $T_{\text{AF}} > 0$ take into account the presence of short-range antiferromagnetic interactions, and $N_0$ is the cation concentration. Important effects of spin-orbit interactions, carrier-carrier correlation,  and mixing between anion and cation wave functions are incorporated into an effective $p-d$ exchange integral $\beta_{\text{eff}}$. In order to evaluate its magnitude we take Mn spin $S = 5/2$ and DOS of holes  at the Fermi level $\rho_{\text{F}}$ from specific heat measurements for Sn$_{1-z}$In$_z$Te with comparable hole densities \cite{Novak:2013_PRB}. The values of  $x_{\text{eff}} =0.024$\ and 0.095 are determined from the Curie constant obtained from our magnetic susceptibility measurements between 100\,K and $T_{\text{Curie}}$; $T_{\text{AF}} = 8xS(S+1){\cal{J}}$, where ${\cal{J}} \simeq 1$\,K (Ref.\,\onlinecite{Gorska:1988_PRB}).  For the experimental magnitudes of $T_{\text{Curie}}$ we then obtain $N_0|\beta_{\text{eff}}| = 0.25 \pm 0.02$\,eV,  a value at the upper bound of those determined for Mn-doped lead chalcogenides \cite{Dietl:1994_PRB}.

\subsection{Search for superconducting precipitates by SQUID magnetometry}

\begin{figure}[tb]
	\includegraphics[width=7cm]{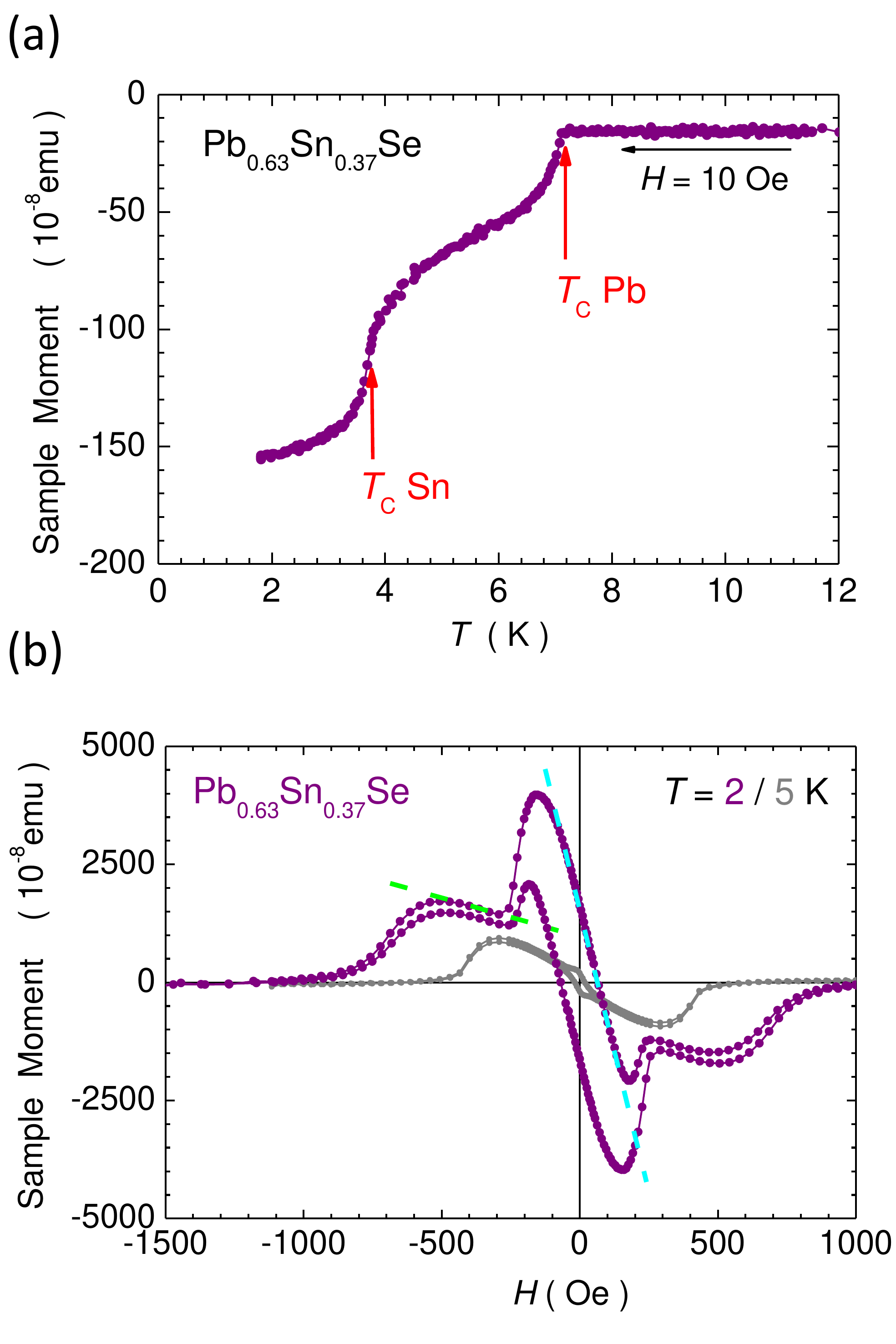}
	\caption{(a) Magnetic moment of a reference sample Pb$_{0.63}$Sn$_{0.37}$Se  measured on cooling in 10\,Oe. Lead and tin superconducting transitions are marked by arrows. (b) Magnetization loops for the same sample after subtracting the diamagnetic component linear in the magnetic field. Dashed lines show slopes taken for evaluation of Pb and Sn masses. The evaluated weight fraction of superconducting precipitates is 0.04\%.}
\label{fig:MSe_T_H}
\end{figure}

We assess the presence of superconducting Pb, Sn, and related precipitates in samples of topological Pb$_{0.20}$Sn$_{0.80}$Te and non-topological Pb$_{0.80}$Sn$_{0.20}$Te, discussed in the main body of the paper, by SQUID magnetometry.
We start by discussing results of magnetization measurements for a reference sample Pb$_{0.63}$Sn$_{0.37}$Se. As shown in Fig.\,\ref{fig:MSe_T_H}, two clear diamagnetic steps in the temperature and magnetic field dependencies of magnetization are recorded, quite accurately corresponding to Pb and Sn superconducting transition temperatures 7.2 and 3.7\,K as well as critical magnetic fields at 2\,K, 700 and 250\,Oe, respectively. The presence of magnetic hystereses points to some pinning. Assuming a spherical shape of precipitates, the field derivative of the magnetic moment $m$ assumes the form,
\begin{equation}
dm_i/d H = -3 \mathcal{M}_i/8\pi\rho_i,
\end{equation}
where  $\mathcal{M}_i$, $\rho_i$, and  $d m_i/d H$ are mass, density, and the experimentally established slopes corresponding to $i =$ Pb or $\beta$-Sn. From the experimental slopes $d{\mathrm{m}}_i/d H$, depicted in Fig.\,\ref{fig:MSe_T_H}(b), we obtain precipitation masses of $\mathcal{M}_{\text{Pb}} = 1.7$\,$\mu$g and  $\mathcal{M}_{\text{Sn}} = 14$\,$\mu$g, which lead to the weight fraction of precipitates $f = 0.04$\%, as sample weight is 38\,mg.

\begin{figure}[tb]
\includegraphics[width=7cm]{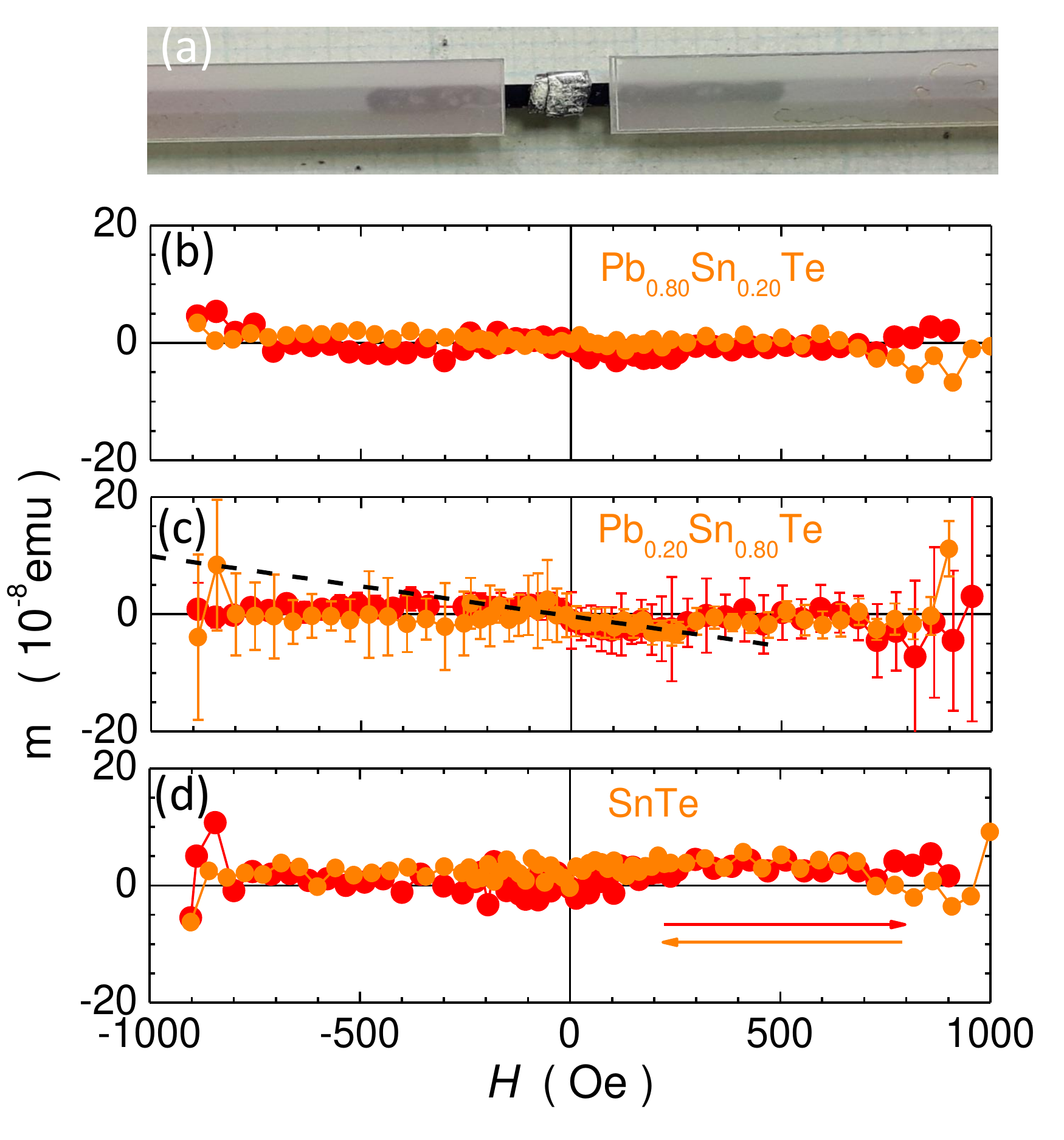}
	\caption{Search for superconducting precipitates at 2\,K in our samples. (a) SQUID sample holder. (b--d) Magnetic moment of Pb$_{1-y}$Sn$_{y}$Te samples after compensating the diamagnetic signal linear in the magnetic field, $y = 0.20, 0.80$, and 1, respectively. Brighter and darker experimental points correspond to measurements for two sweeping directions of the magnetic field. The evaluated upper limit of the weight fraction of superconducting precipitates is 0.1\,ppm [dashed line in (c)].}
\label{fig:MTe_H}
\end{figure}

\begin{figure}[tb]
	\includegraphics[width=7cm]{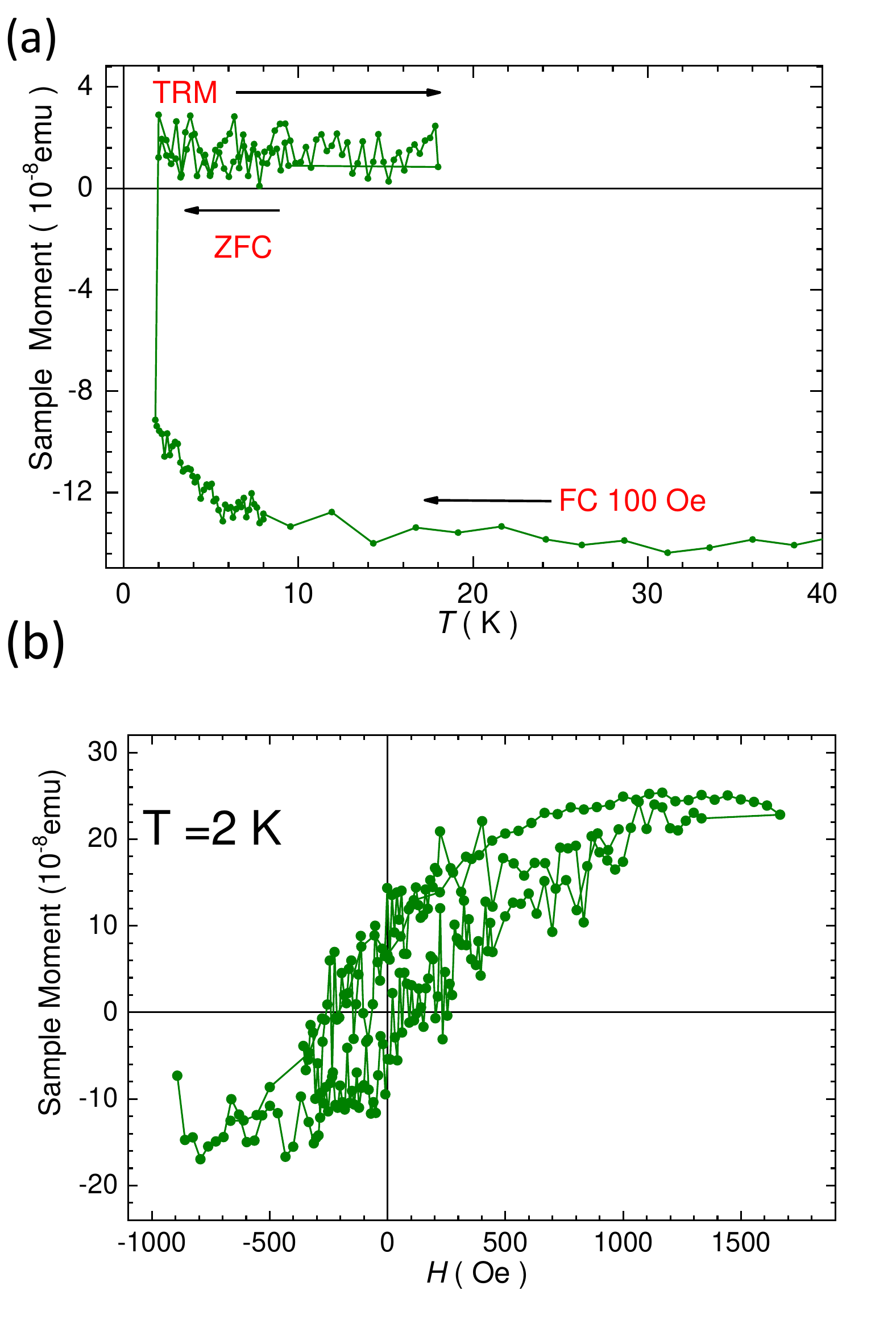}
	\caption{Search for superconducting precipitates in silver paint employed for soft point contact spectroscopy deposited on Si. (a) Magnetic moment as a function of on temperature. (b) Magnetization vs. magnetic field at 2\,K. Data are corrected for substrate diamagnetism.}
	\label{fig:MTe_H_Ag}
\end{figure}

Neither Pb$_{0.20}$Sn$_{0.80}$Te nor Pb$_{0.80}$Sn$_{0.20}$Te shows such a signal. We assign a large concentration of Sn and Pb precipitates in Pb$_{0.63}$Sn$_{0.37}$Se to the fact that Sn content $y = 0.37$ is close the solubility limit $y \simeq 0.40$. In order to test our telluride samples Pb$_{1-y}$Sn$_{y}$Te with even higher sensitivity and, in particular, to compensate a relatively large bulk diamagnetic signal, a dedicated sample holder has been prepared. As illustrated in Fig.\,\ref{fig:MTe_H}(a), 20--40\,mm long and 5\,mm wide strips of sapphire are glued to the silicon sample holder \cite{Gas:2019_MST} forming in the holder center a gap of the length of about 5\,mm, to which studied samples are inserted [Fig.\,\ref{fig:MTe_H}(a)]. As these sapphire strips extend from the gap for about 8\,cm each way, by adjusting the mass of measured samples to be inserted into the gap, the total signal can be made field independent (typically down to 2\% of the initial slope), provided that both the sample and the sapphire stripe do not contain any magnetic or superconducting inclusions. We use a pure GaAs sample to determine a background signal of the sample holder.  By adjusting masses of Pb$_{1-y}$Sn$_{y}$Te samples of interest here, and after correcting for sapphire response$^{S5}$, we obtain SQUID signals presented in Fig.\,\ref{fig:MTe_H}(b--d). The magnitude of noise indicates that for a mean mass of our samples, i.e., 60\,mg, the upper limit of the weight fraction of precipitates that could give a response of the type presented in Fig.\,\ref{fig:MTe_H} is 0.1\,ppm.

\begin{figure*}[tb]
	\includegraphics[width=15cm]{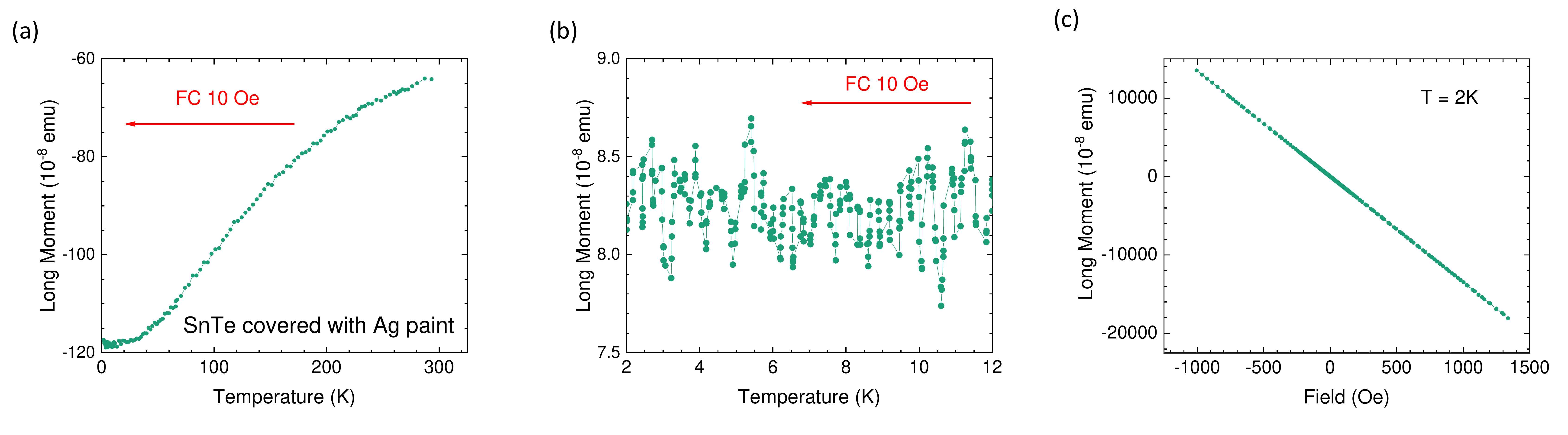}
	\caption{Search for a magnetic indication of superconductivity in SnTe sample covered entirely by silver paint in an in-plane magnetic field. (a,b) Magnetic moment vs. temperature measured on cooling in a magnetic field of 10\,Oe. (c) Magnetization vs. magnetic field at 2\,K showing diamagnetism of SnTe.}
	\label{fig:SnTe_Ag}
\end{figure*}

It can be hypothesized that silver paint employed for point contact spectroscopy contains superconducting precipitates which might account for the observed features in differential conductance. This scenario is excluded by magnetization measurements of silver paint deposited on a Si sample presented in Fig.\,\ref{fig:MTe_H_Ag}. A signal from residual paramagnetic impurities superimposed on a temperature independent diamagnetic response is found. Furthermore, considering a possible diffusion of Sn or Pb towards the Ag electrode or the dislocation formation by the temperature stress, the magnetic response of the SnTe sample with the topological surface covered entirely by silver paint has been measured by SQUID magnetometry. As shown in Fig.\,\ref{fig:SnTe_Ag}, no superconductive features are detected down to the lowest accessible temperature of 2\,K.

\begin{figure}[tb]
\includegraphics[width=9.5cm]{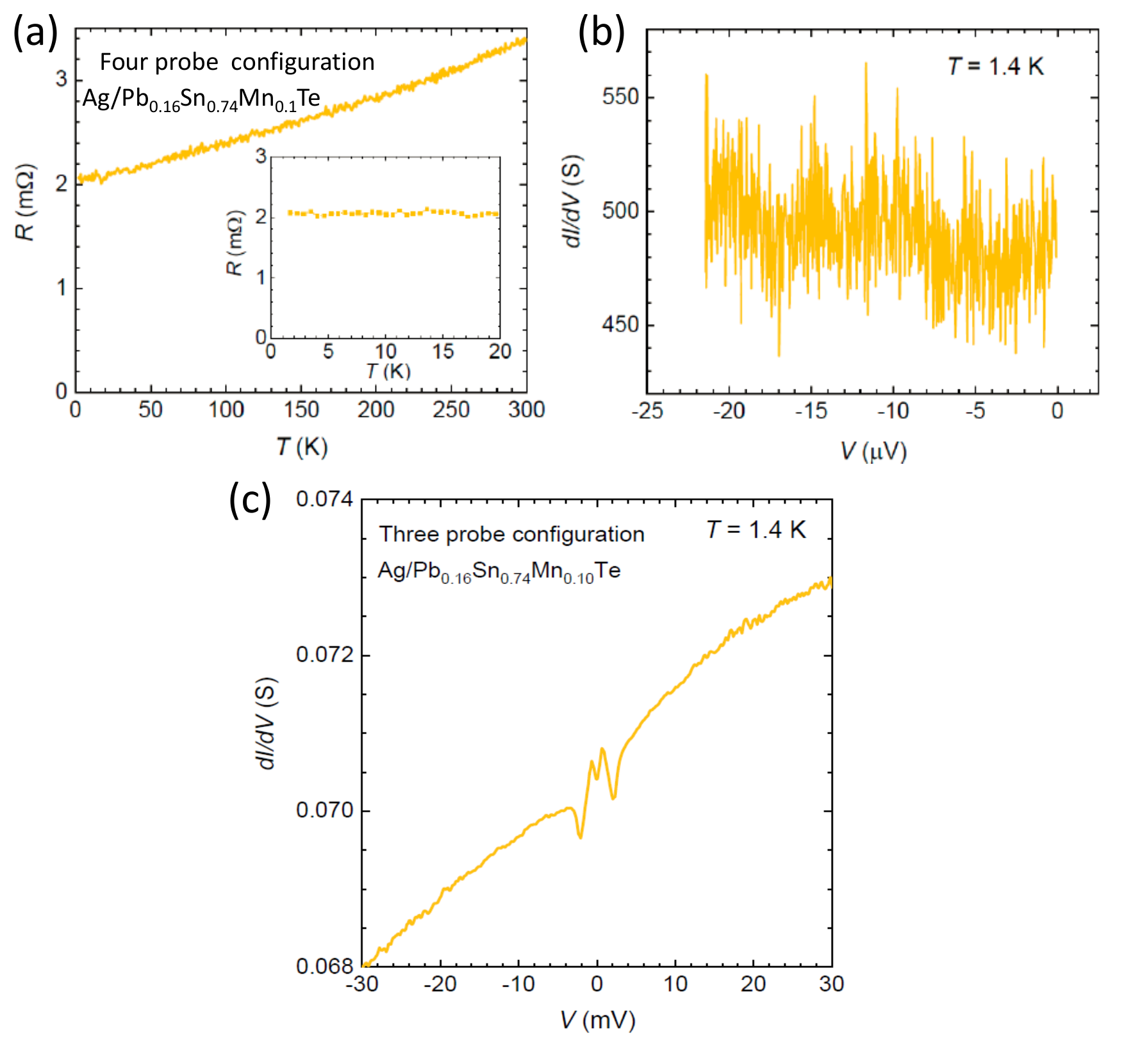}
	\caption{Search for resistive indication of superconductivity at Ag/TCI interface.
 Resistance $R$ of a bilayer of an Ag film deposited onto Pb$_{0.16}$Sn$_{0.74}$Mn$_{0.10}$Te measured by a four probe method as a function of temperature (a) and bias voltage (b). Inset presents the low-temperature range of the $R(T)$ dependence. No superconductivity is detected down 1.4\,K. (c) Differential conductance measured in a three probe configuration showing a zero-bias conductance peak measured across the soft-point contact at the Ag/TCI interface.}
\label{fig:AgTCI}
\end{figure}

\subsection{Resistance and differential conductance of Ag/TCI and Au/TCI interfaces}

\begin{figure}[tb]
	\includegraphics[width=7cm]{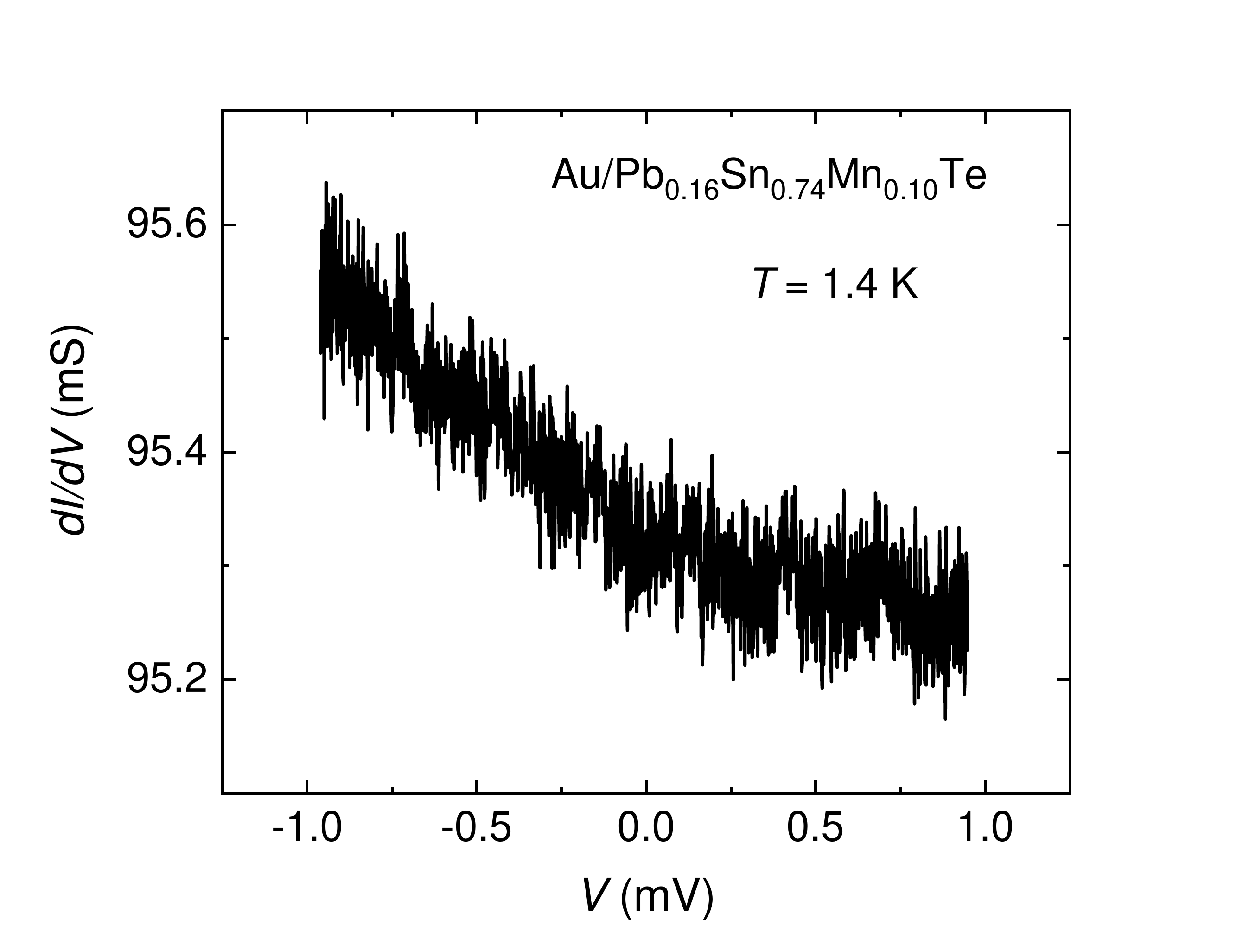}
	\caption{Differential conductance of Au welded contacts to TCI. Spectrum does not show any signatures of zero-bias anomaly or a gap.}
	\label{fig:AuTCI}
\end{figure}

Our point-contact spectroscopy experiments have been performed by using silver paint. The remaining two contacts are spot-welded into the sample to ensure a good electrical connection.  To check properties of the Ag/TCI interface we have deposited a 50\,nm layer of Ag using ultra high vacuum e-beam evaporator onto a ferromagnetic Pb$_{0.16}$Sn$_{0.74}$Mn$_{0.10}$Te sample.  Resistance measurements of the Ag/TCI bilayer have been carried out employing a four probe configuration with the current probes spot-welded to the sample and the voltage probes fixed by silver paint to the silver layer. The resistance measurements as a function of temperature [Fig.\,\ref{fig:AgTCI}(a)] and bias voltage [Fig.\,\ref{fig:AgTCI}(b)] do not exhibit any noticeable superconductivity in the studied temperature and bias voltage ranges. At the same time, in the three contact geometry, with one of the silver paint contacts serving as a current and voltage probe, we detect a ZBCP, as shown in Fig.\,\ref{fig:AgTCI}(c). Thus we conclude  that the presence of the ZBCP is not accompanied by a superconductivity of the Ag/TCI interface. Furthermore, a point contact spectroscopy experiment is performed by using only spot-welded contacts. No features in differential conductance are found, as  show in Fig.\,\ref{fig:AuTCI}. This indicates that no ZBCP appears if the interface is damaged by the welding process.

\subsection{Additional data on point contact spectroscopy}
\begin{figure}[tb]
	\includegraphics[width=9.5cm]{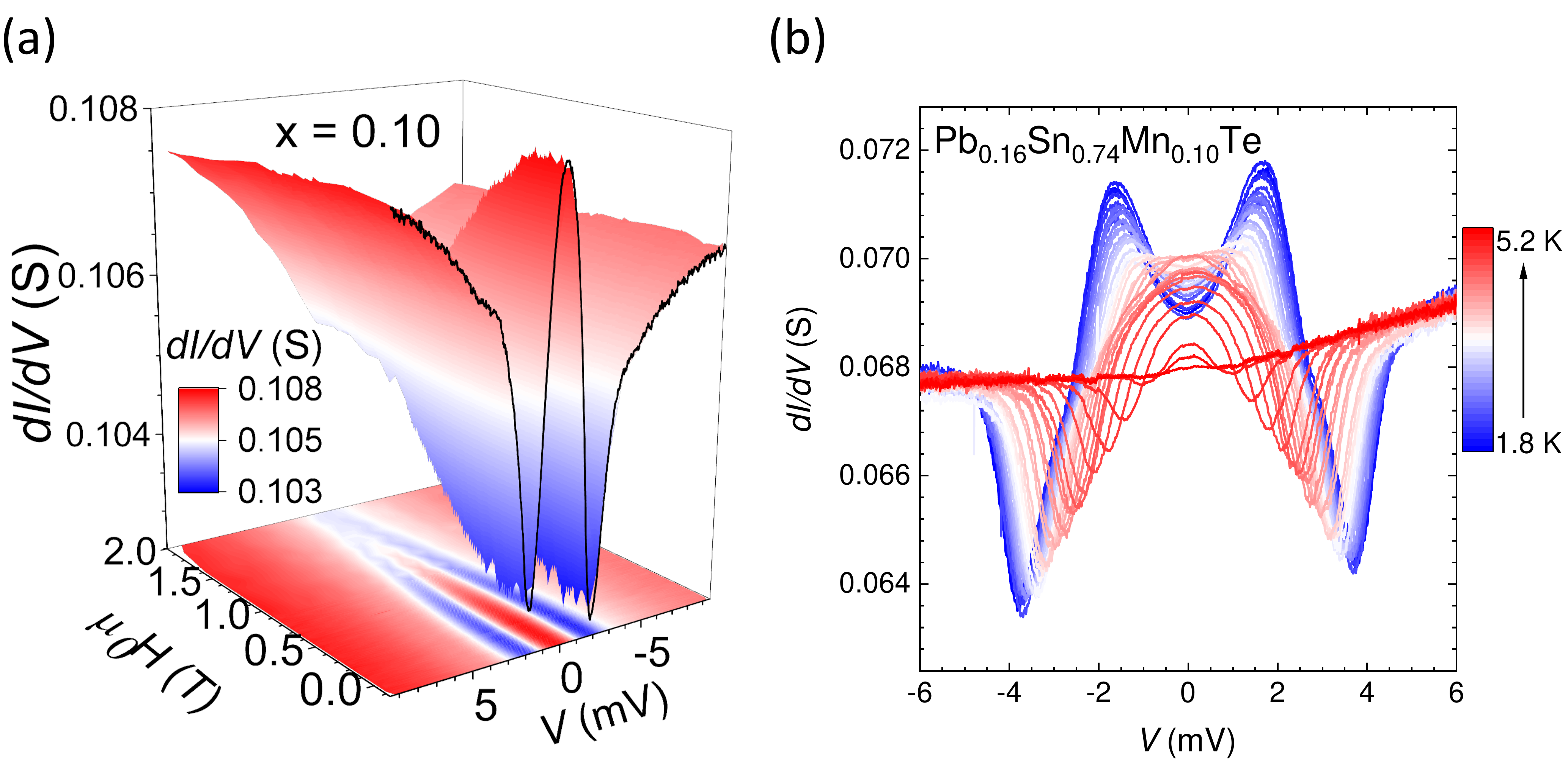}
	\caption{(a) Evolution of the spectra for Pb$_{0.16}$Sn$_{0.74}$Mn$_{0.10}$Te as a function of magnetic field in the thermal regime. (b) Temperature evolution of conductance spectra for Pb$_{0.16}$Sn$_{0.74}$Mn$_{0.10}$Te in the intermediate regime.}
	\label{fig:PCS_Additional}
\end{figure}

\begin{figure}[tb]
	\includegraphics[width=9cm]{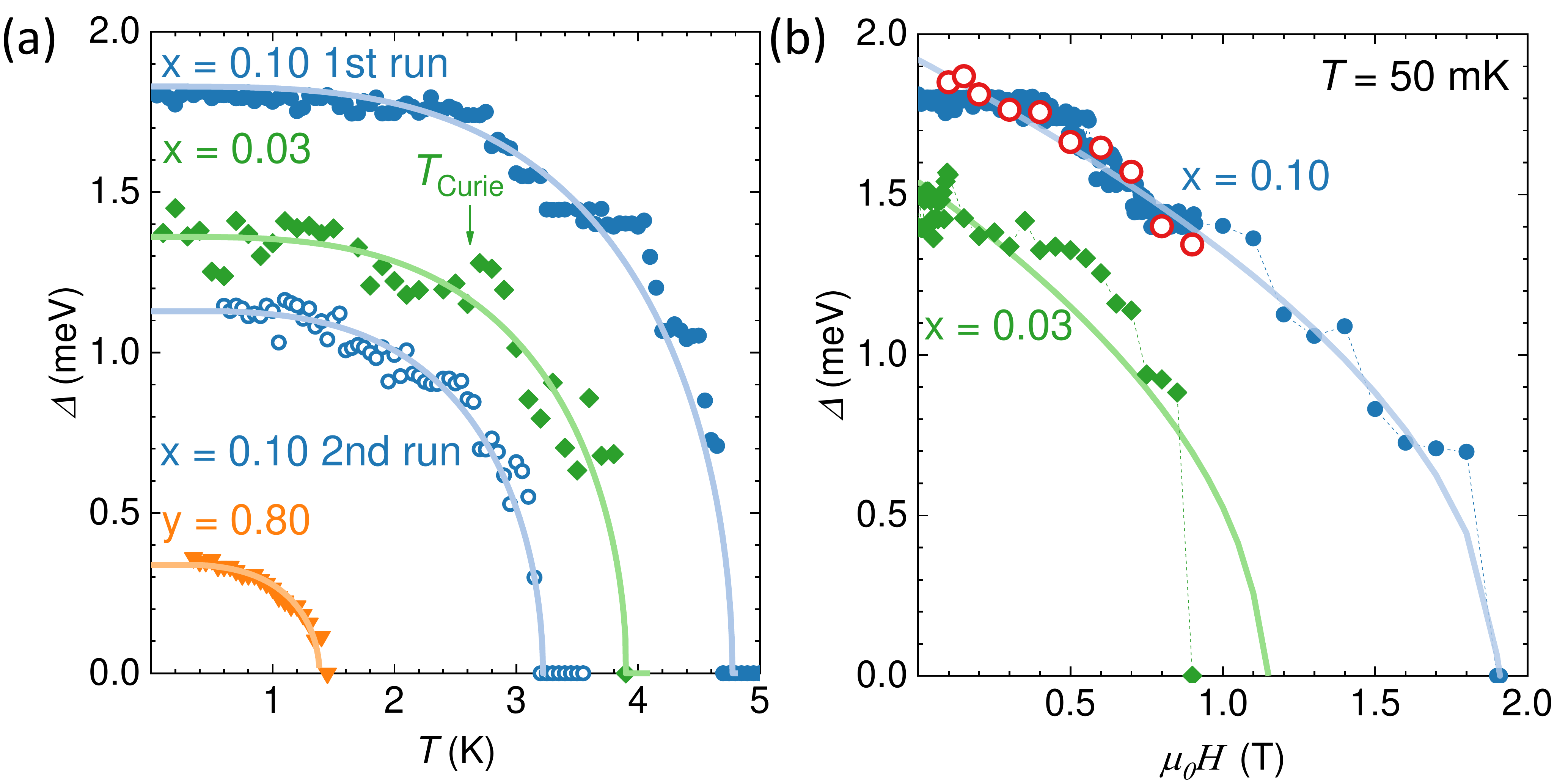}
	\caption{ Gap dependence $\Delta$ on temperature (a) and magnetic field (b)  evaluated from the conductance minima from the differential conductance spectra presented in Figs.1 and 2 in the main manuscript. The red open circles on the panel (b) correspond to in-plane applied magnetic field.}
	\label{fig:Gap_Additional}
\end{figure}

\begin{figure}[tb]
	\includegraphics[width=6.5cm]{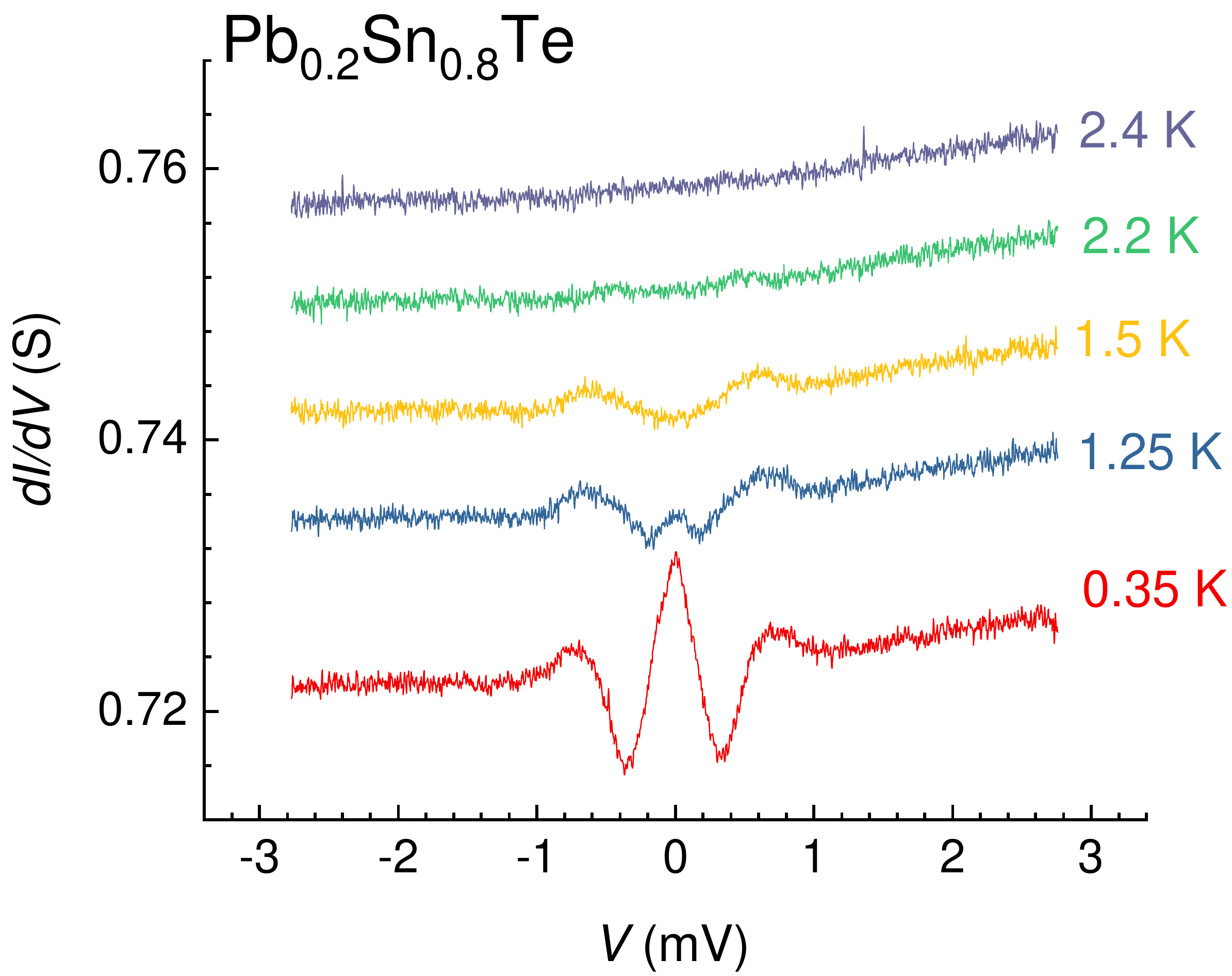}
	\caption{The figure illustrates slices of Fig.1(b) from the main manuscript. It shows how ZBCP vanish as a function of temperature leaving clear gap structure. At 2.4 K spectrum is featureless.}
	\label{fig:GapEst-PbSnTe}
\end{figure}

\begin{figure}[tb]
	\includegraphics[width=6.5cm]{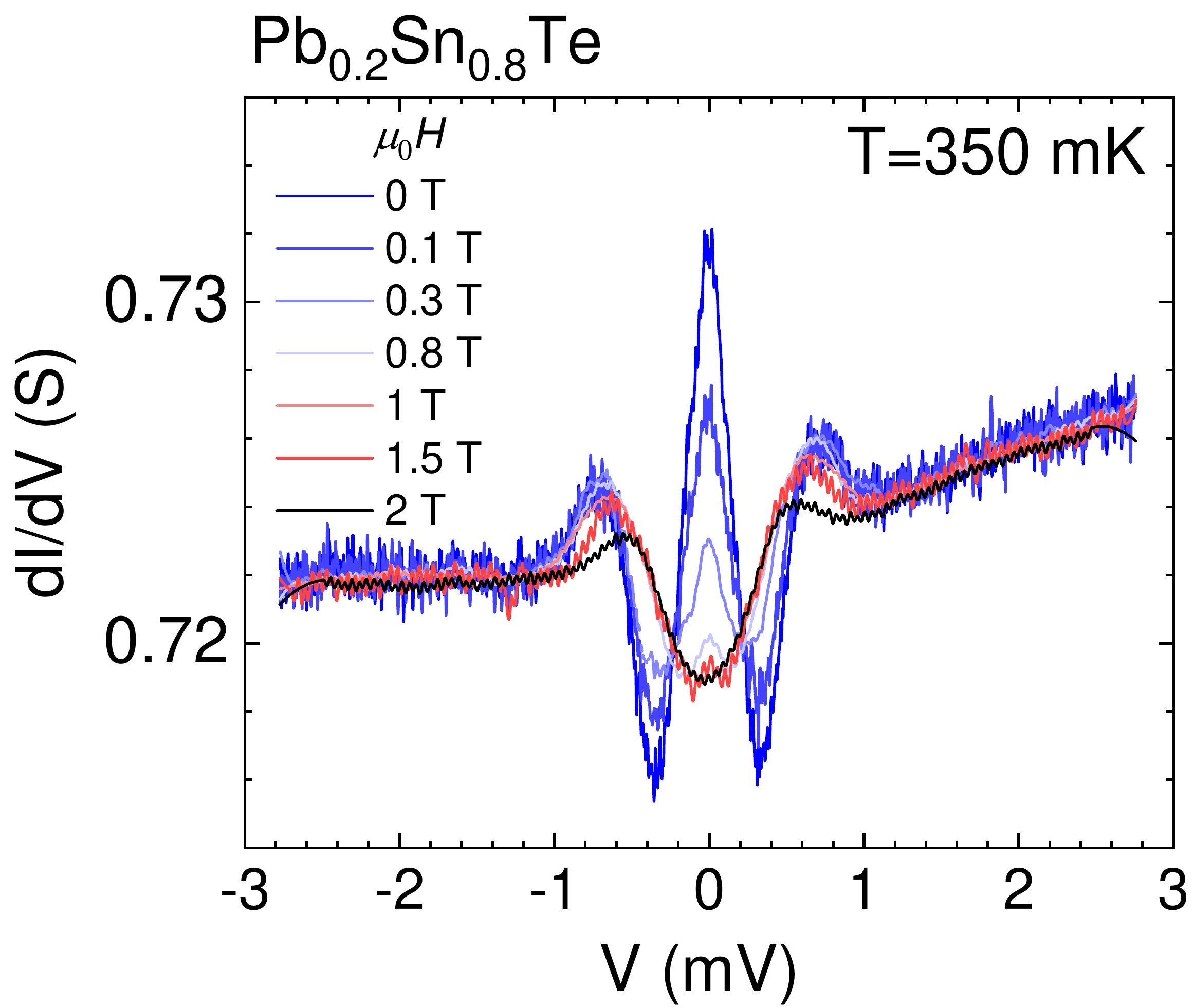}
	\caption{ZBCP quenched as a function of magnetic field in point contact spectroscopy of Pb$_{0.2}$Sn$_{0.8}$Te.}
	\label{fig:PbSnTeHSpectrum}
\end{figure}

Figure~\ref{fig:PCS_Additional} supplements data presented in the Fig.~2 in the main text. Figure~\ref{fig:Gap_Additional} presents dependence of the gap, interpreted as a differential conductance minimum, on temperature (a) and the magnetic field (b) in the thermal regime. This is to support the data presented in the Figs.~3(b,c) in the main text. Figure~\ref{fig:GapEst-PbSnTe} shows how spectroscopic features vanish as a function of temperature. First, the zero-bias peak decreases as a function of temperature, leaving clear gapped spectrum at 1.5\,K. Similarly, the magnetic field quenches zero-bias peak as depicted in Fig.~\ref{fig:PbSnTeHSpectrum}.

\begin{figure}[tb]
	\includegraphics[width=9.5cm]{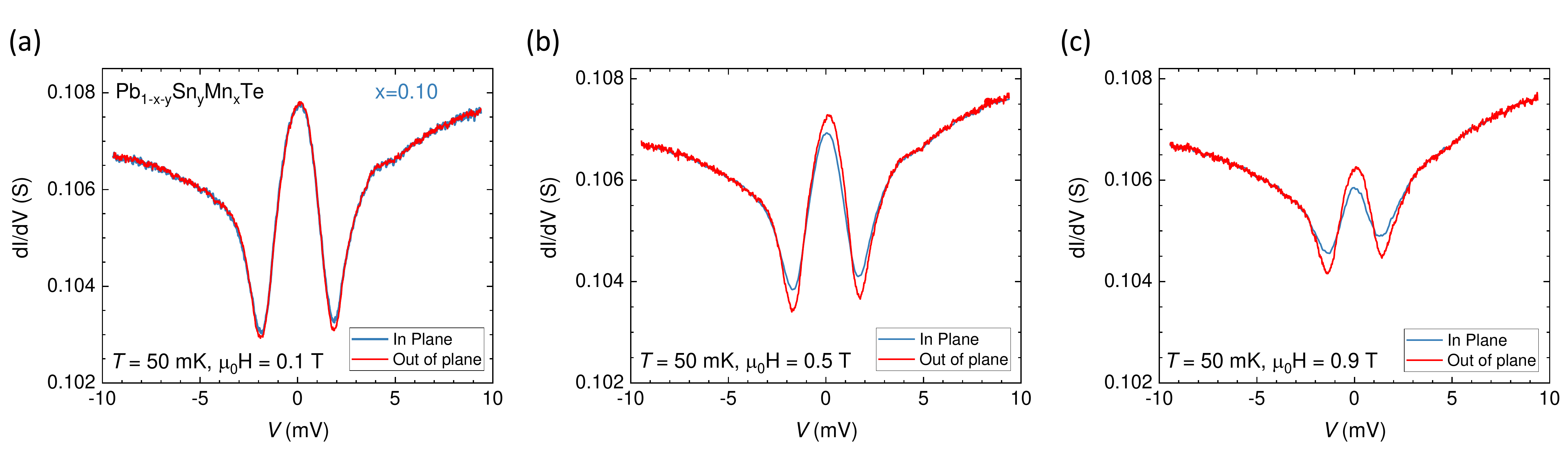}
	\caption{Differential conductance spectra at various perpendicular and in-plane magnetic fields at 50\,mK for Pb$_{0.16}$Sn$_{0.74}$Mn$_{0.10}$Te sample.}
	\label{fig:AMR}
\end{figure}

\subsection{Magnetic anisotropy of differential conductance}
In order to probe magnetic anisotropy of the point-contact spectra,  differential conductance of the Pb$_{0.16}$Sn$_{0.74}$Mn$_{0.10}$Te sample has been measured for various fields perpendicular and parallel to the sample surface. As seen in Fig.\,\ref{fig:AMR} there is some anisotropy in the peak heights but no noticeable differences in the width of the spectra. A minor magnitude of magnetic anisotropy reemphasizes that the observed low temperature phase has no global 2D character.

\begin{figure}[tb]
	\includegraphics[width=9.5cm]{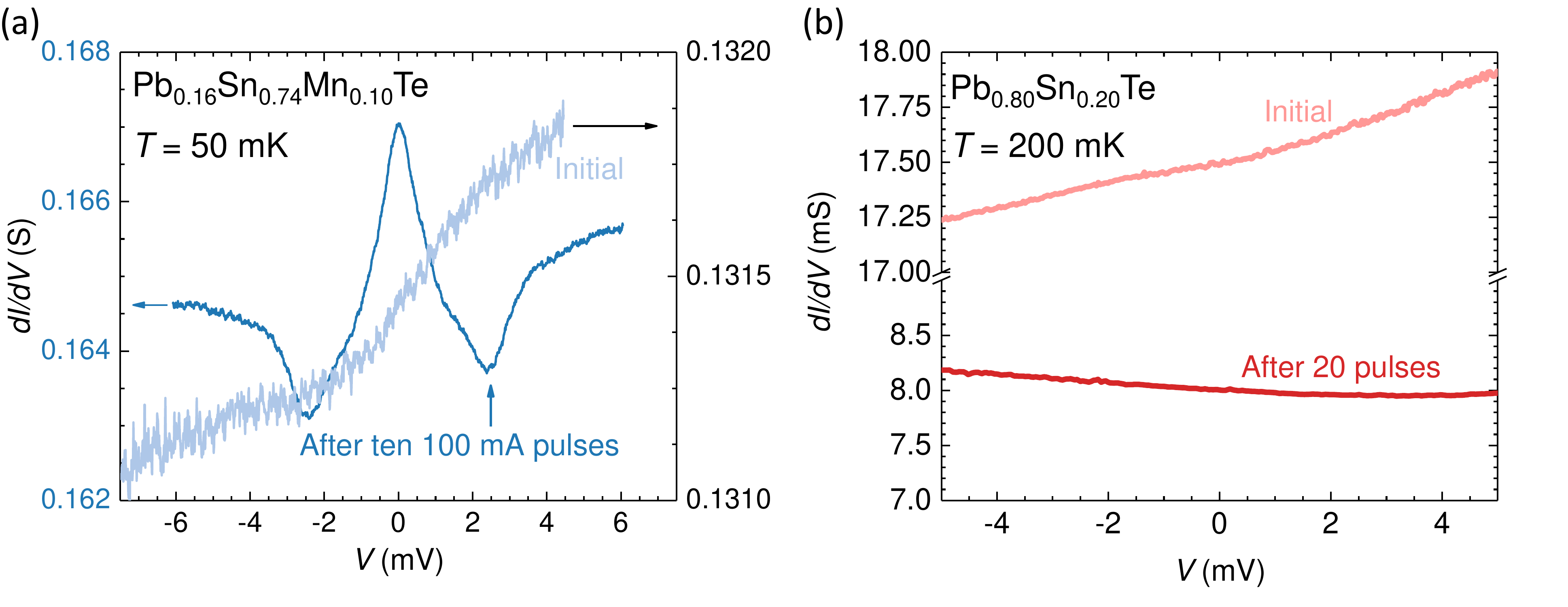}
	\caption{Effect of current pulses on differential conductance characteristics in  topological and non-topological samples. (a) In topological samples pulses can restore a peak structure in the case of contacts that have been initially featureless. (b) Differential conductance remains featureless in non-topological samples.}
\label{fig:current_pulses}
\end{figure}

\subsection{Effect of current pulses}
In most cases freshly prepared silver paint contacts reveal zero-bias conductance peak in topologically non-trivial samples. However, some of the contacts stopped working after temperature sweeps or showed no structure during initial measurements. We have found, as could be expected for phenomena relying on properties of surface steps, that characteristics of differential conductance spectra in topological samples can be modified by current pulses, as shown in Fig.\,\ref{fig:current_pulses}(a). A typical pulse sequence consists of 1\,ms pulses with amplitude 100\,mA, with the repetition rate of 10\,Hz. Such a sequence does not awake any structure in conductance of non-topological Pb$_{0.80}$Sn$_{0.20}$Te [Fig.\,\ref{fig:current_pulses}(b)].

\begin{figure}[tb]
	\includegraphics[width=9.5cm]{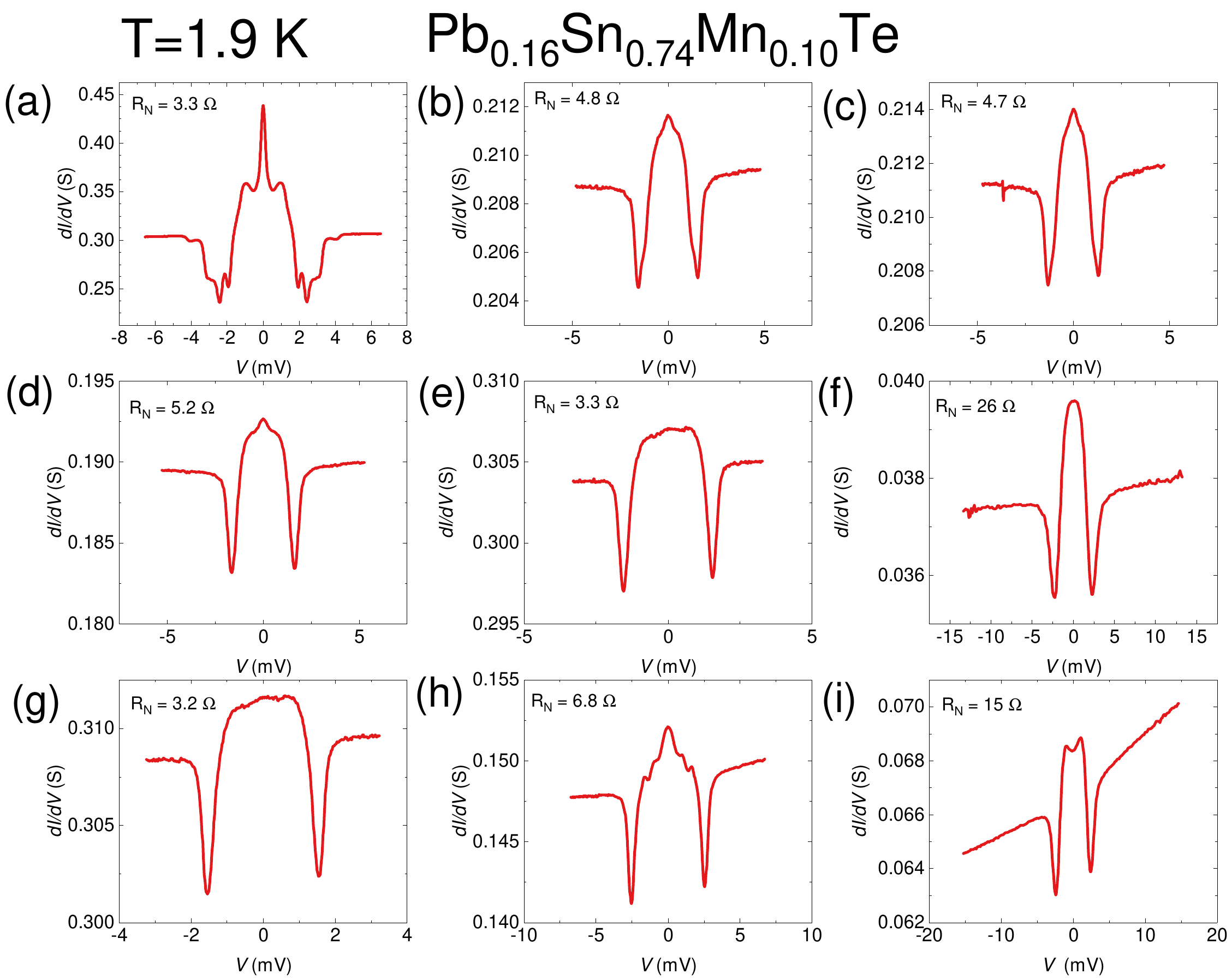}
	\caption{Various point contact spectra obtained by applying small 1\,ms current pulses across the contacts for Pb$_{0.16}$Sn$_{0.74}$Mn$_{0.10}$Te. The pulse height varied from 10 mA to 50 mA depending on the contact resistance.}
	\label{fig:current_pulses2}
\end{figure}

\subsection{AFM scans after point-contact spectroscopy studies}

\begin{figure}[tb]
	\includegraphics[width=7cm]{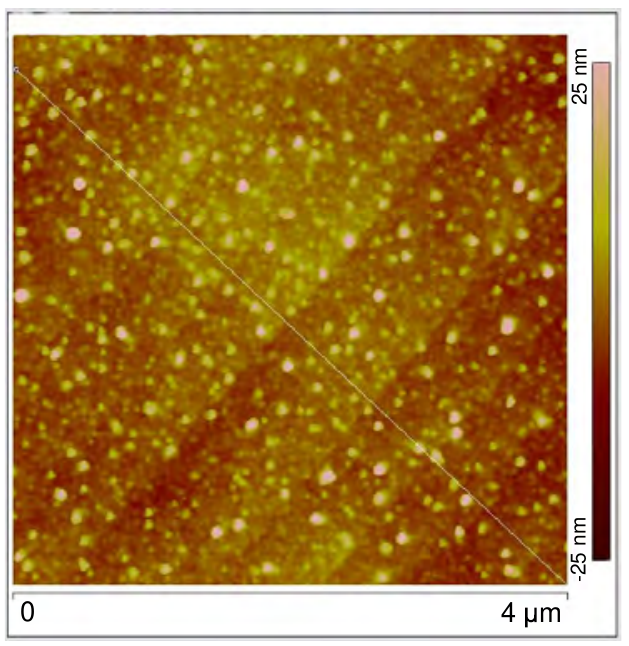}
	\caption{AFM image of wire saw cut (011) surface of the Pb$_{0.16}$Sn$_{0.74}$Mn$_{0.10}$Te single crystal after series of etchings and measurements. This surface shows multilayer steps and a relatively high roughness.}
	\label{fig:AFM_Meas}
\end{figure}

Figure ~\ref{fig:AFM_Meas} presents the AFM image of the (011) surface obtained by a wire-saw cut and subject to a series of etchings. As seen surface roughness are significantly larger compared to the freshly grown or cleaved surfaces, whose AFM images are presented in Fig.~4 of the main text.

\ \\
\vspace{-3cm}
\ \\

%



\clearpage

\end{document}